\documentclass[12pt,a4paper,twoside,openright]{book}
\usepackage{a4wide}
\usepackage{amsmath,amsfonts,amssymb,latexsym}
\usepackage{epic,eepic}
\usepackage[cp850]{inputenc}
\usepackage{setspace}
\usepackage{graphicx}
\usepackage{epsfig}
\usepackage[dvips]{lscape}
\usepackage{fancyhdr}
\renewcommand{\headrulewidth}{0.4pt}
\newcommand{\chshort}{} 
\newcommand{\Date}{}
\newcommand{\Revision}{}
\newcommand{\chset}[1]{\renewcommand{\chshort}{\thechapter. #1}}
\newcommand{\chsets}[1]{\renewcommand{\chshort}{#1}}
\makeatletter
\def\cleardoublepage{\clearpage\if@twoside \ifodd\c@page\else
 \hbox{}
 \vspace*{\fill}
 \thispagestyle{empty}
 \newpage\fi\fi}
\makeatother
\usepackage{cite}
\usepackage[small]{caption}
\def \be{\begin{equation}}
\def \ber{\begin{eqnarray}}
\def \ee{\end{equation}}
\def \eer{\end{eqnarray}}

\begin{document}
\pagenumbering{roman}
\fancyhf{} 
\fancyhead[CE]{M. H Szymanska --- Bose condensation and lasing in optical microstructures}
\fancyhead[LO]{\sl \leftmark}
\fancyfoot[CO,CE]{\thepage}
\fancypagestyle{plain}{
\renewcommand{\headrulewidth}{0pt}
\fancyhf{}
\fancyfoot[C]{\thepage}
\fancyfoot[L]{\Date}
\fancyfoot[R]{\Revision}}
\thispagestyle{plain}
\frontmatter
\begin{center}
\Large
\vspace*{\stretch{1}}
\Huge
\textbf{Bose Condensation and Lasing\\
        in Optical Microstructures}

\vspace*{\stretch{1}}

\textbf{ Part 1}  \\[.5cm]

\vspace*{\stretch{1}}
\normalsize

\large

\textbf{ Marzena Hanna Szymanska}  \\[.5cm]

\normalsize
 Trinity College \\
University of Cambridge

\vspace*{\stretch{1}}
\normalsize

\begin{center}
\includegraphics[width=0.16\linewidth,angle=0]{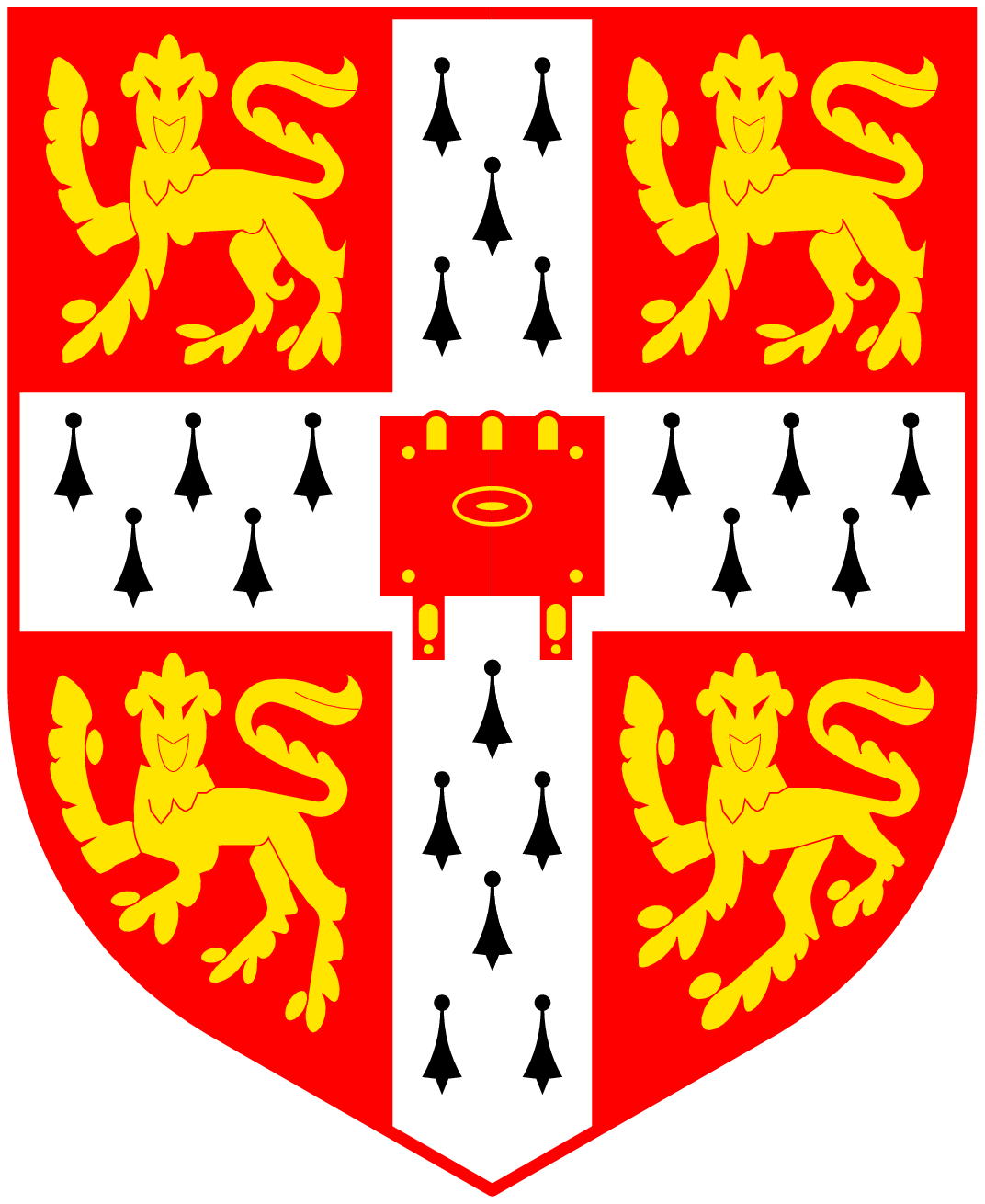}
\end{center}

\vspace*{\stretch{1}}
\normalsize

Dissertation submitted for the degree \\
of Doctor of Philosophy at the \\
University of Cambridge
\\[.2cm] 
{\em October 2001}

\end{center}

\tableofcontents
\chapter*{Preface}
\addcontentsline{toc}{chapter}{Preface}
\chset{Preface}

This dissertation describes work undertaken at the Cavendish
Laboratory between October 1998 and October 2001, under the
supervision of Professor Peter Littlewood.

Some of the work presented in this thesis has been published, has been
submitted for publication, or is in preparation for publication, in
the papers:

\begin{itemize}
\item M.~H.~Szymanska, P.~B.~Littlewood, R.~J.~Needs, ``Excitons in T-shaped
quantum wires'', {\it Phys.\ Rev.\ B}, 63, 205317 (2001). 

\item J.~Rubio, L.~N.~Pfeiffer, M.~H.~Szymanska, A.~Pinczuk, S.~He,
H.~U.~Baranger, P.~B.~Littlewood, K.~W.~West, B.~S.~Dennis,
``Coexistence of Excitonic Lasing with Electron-Hole Plasma,
Spontaneous Emission in One-Dimensional Semiconductor Structures
'', accepted for publication in {\it Solid State Commun.}.
 
\item M. H. Szymanska, P. B. Littlewood, ``Bose Einstein Condensation
of cavity polaritons and lasing - crossover between the two regimes'',
in preparation for {\it Phys.\ Rev.\ Letters}.

\item J.~Rubio, L.~N.~Pfeiffer, M.~H.~Szymanska, A.~Pinczuk, S.~He,
H.~U.~Baranger, P.~B.~Littlewood, K.~W.~West, B.~S.~Dennis, ``Two-mode
lasing in a quantum wire'' in preparation for {\it Phys.\ Rev.\ B}
\end{itemize}

Except where stated otherwise, this dissertation is the result of my
own work and contains nothing which is the outcome of work done in
collaboration. This dissertation has not been submitted, in whole or
in part, for any degree or diploma other than that of Doctor of
Philosophy at the University of Cambridge.

\begin{flushright}
Marzena Hanna Szymanska \\ Trinity College \\ October 2001
\end{flushright}

\chapter*{Summary}
\addcontentsline{toc}{chapter}{Summary}
\chset{Summary}

There are two distinct phenomena where quantum coherence has been
observed on the macroscopic scale: the laser and Bose Einstein 
condensation. Although the physics underlying these processes is
fundamentally different, they both could potentially be observed in
the same microcavities, and both would be a source of a coherent
light.

In the first part of this thesis I study the intermediate regime
between ordinary lasing and a BEC of exciton polaritons. I take into
account the fermionic structure of polaritons, treating the excitons
as two-level systems coupled to a single mode in a microcavity. I
introduce decoherence and dissipation processes to this system.
Employing many-body Green function techniques, similar to those
used by Abrikosov and Gor'kov in their theory of gapless
superconductivity, I provide a mathematical structure that unifies
models of lasers with models of condensates. This allows me to study
the stability of the polariton condensate with respect to decoherence
processes and the crossover between the polariton condensate and the
laser.  I give detailed indications of a regime in which the
condensate should be observed to guide experimental work and show
how to distinguish the Bose condensate from a laser.

The second part of this thesis is concerned with properties of
excitons and modelling of excitonic lasing in quasi-one-dimensional
quantum wires.  I develop a very general numerical method of
calculating the properties of wires with different shapes and
materials. Using this method I study the properties of very wide range
of T-shaped quantum wires. I also develop an analytical model for
the two-mode laser. Using my theory I explain the detailed issues raised
by experiments. The method is being used in designing a high
performance 1D wire laser in collaboration with Bell Laboratories.

\chapter*{Acknowledgements}
\addcontentsline{toc}{chapter}{Acknowledgements}
\chset{Acknowledgements}

I am deeply indebted to my supervisor, Peter Littlewood for
his guidance, support and encouragement, for his constant optimism
over the last three years and for his advice in many matters of my
academic life.
 
I owe particular thanks to Paul Eastham for many exciting discussions
and his stimulating criticism.  I am very grateful to Richard Needs,
Peter Haynes, Andrew Porter and Greg McMullan for their help and
advice concerning the computational techniques. I am pleased to thank
Hidefumi Akiyama, Loren Pfeiffer and Aron Pinczuk for numerous
interesting discussions during my visits at Bell-Laboratories. I would
like to record my gratitude to Ben Simons and David Khmeltnitskii for
their helpful opinions and suggestions, and Robert Moir, Chris Pickard,
Mike Towler and Jem Pearson for proofreading parts of this thesis.

Special mentions are due to Tracey Ingham, our group secretary, for
being so helpful in solving numerous problems of everyday life in TCM.
I also owe particular thanks to Yong Mao, Mercedes Hinton and
Francesca Marchetti for their support during the ups and downs of my
life in Cambridge.  Finally and most importantly I would like to thank
my family for their love, help and encouragement throughout my PhD.

I acknowledge financial support from Lucent Technologies, Trinity
College Cambridge, Cambridge Overseas Trust and an ORS award.

\chapter*{Introduction}
\addcontentsline{toc}{chapter}{Introduction}
\chsets{Introduction}

Coherence is a property of waves which describes how much they
resemble a smooth sine wave. In a quantum theory matter, similarly to
light, has a wave-like character and can exhibit coherence. However,
neither the wave nature of matter nor coherent states are generally
observed in everyday life. This is because, usually, only a tiny
fraction of particles occupy the same quantum state leading to an
incoherent, classical world. 

It is, however, possible to produce systems in which a large number of
particles occupy the same coherent state observable on a macroscopic
scale. One of the most common and widely used for practical
applications is the laser - a coherent state of light. Observation of
a coherent state of massive, interacting and number conserved bosons,
known as a Bose Einstein condensate (BEC), is very difficult due to
large decoherence effects, and requires very low temperatures. Liquid
Helium and trapped atomic gases remain the only systems
where BEC is experimentally observed and controlled. Superconductivity
- a coherent state of paired electrons - closes the list of current
experimental realisations of macroscopic quantum coherent
states. These few experimental realisations have had a large impact on
an understanding of quantum physics.

Another candidate for BEC, which would be very interesting for
practical applications, is excitons - bound states of electrons and
holes which occur in semiconductors. Together with the decoherence
effects the major difficulty in obtaining the coherent state of
excitons is their non-equilibrium nature. They decay into
light. Excitons can, however, be placed in an optical cavity, where
photons are prevented from escaping. If the exciton-photon interaction
is sufficiently strong they form new particles: cavity
exciton-polaritons which are bosons and so can Bose condense.

Miniaturisation and improvements in the quality of optical cavities in
recent years bring new possibilities for the observation of coherent
effects. In cavities the size of a wavelength of light, called
microcavities, the exciton-photon interaction is particularly strong
and the decoherence is reduced. The experimental observation of Bose
condensation of cavity polaritons might become possible. Such a
polariton condensate could also be a new source of coherent light
where, in contrast to lasers, no population inversion would be
required.

Although the first laser action was observed long ago, extensive
research is still being done in the search for miniaturised,
low-threshold lasers. Recent achievements in manufacturing good
quality low-dimensional microstructures provide the possibility for
much better performance of optical devices.

My thesis addresses two aspects of this subject. In the first part we
develop a theory which connects Bose-Einstein condensation of
microcavity polaritons and lasing. The second part is devoted to
modelling excitonic lasing in quasi-one-dimensional, T-shaped quantum
wires.

\section*{Bose Einstein condensation of microcavity polaritons and
lasing crossover between the two regimes}

There are two distinct phenomena where quantum coherence has been
observed on the macroscopic scale: the laser and Bose Einstein
condensation. Although the physics underlying both processes is
fundamentally different, they both could potentially be observed in
the same microcavities, and both would be a source of a coherent
light.

The BEC of excitons or exciton-polaritons is still an open issue,
undergoing extensive experimental research.
Indications of some excitonic coherence effects in semiconductor
microcavities were reported but whether BEC of exciton-polaritons took
place remains unresolved.

As the BEC of polaritons could be a source of a coherent light the
issue arises how the condensation can be distinguished from an
ordinary lasing. The development of a theory which connects these two
regimes in microcavity, predicts the major signs of a condensate,
and checks its stability with respect to decoherence processes is of both
fundamental interest and will help experimentalists in the
search for condensation.

Generalisation of existing laser theories to include the media
coherence effects seems very promising as a polariton condensate could
be a source of a coherent light without the population inversion.

We study the intermediate regime between ordinary lasing and a BEC of
exciton polaritons. We take into account the fermionic structure of
polaritons, treating the excitons as two-level systems coupled to a
single mode in a microcavity. We introduce decoherence and dissipation
processes to this system.

In the limit of very large decoherence we obtain a conventional laser.
In the opposite regime we can examine the stability of the BEC of
polaritons. 

The work, although motivated by the experiments on semiconductor
excitons, is much more general, and can be applied to any two-level
systems interacting via photons confined in a cavity.

\section*{Excitons in T-shaped quantum wires}

Optical properties of electrons and holes confined to few dimensions
are of particular interest for optical and electronic devices.  As the
dimensionality of the structure is reduced, the density of states
tends to bunch together leading to a singularity in the 1D case. At
the same time the excitonic interaction in 1D is enhanced. For the
ideal 1D quantum wire, the excitonic binding energy and oscillator
strength diverge suggesting that they can be greatly increased in
practice for quasi-1D wires. Both effects provide possibilities for
improving the performance of optical devices such as semiconductor
lasers.

A much studied quasi-1D experimental system is the T-shaped quantum
wire formed at the intersection of two quantum wells. Laser emission
from the lowest exciton state in T-shaped quantum wires has been
experimentally observed, and an interesting observation of two-mode
lasing in such wires has been recently reported.

We develop a rate equation model to understand this two-mode
lasing phenomena. The model gives very good agreement with
experimental data, which suggests that we have lasing from two
different excitonic states in the structure. Thus excited states
seem to be very important for the operation of excitonic lasers.

Motivated by the growing experimental realisation of these structures,
as well as the interesting report of lasing phenomena, we have developed
a numerical method to calculate energies, oscillator strengths for
radiative recombinations, and two-particle wave functions for the
ground state exciton and the excited states in a T-shaped quantum
wire. These calculations are being used to design improved excitonic
lasers which will operate at room temperature.

\cleardoublepage
\mainmatter
\pagenumbering{arabic}
\fancyhf{} 
\fancyhead[CE]{M. H Szymanska --- Bose Condensation and Lasing in Optical Microstructures }
\fancyhead[LO]{\chshort}
\fancyhead[RO,LE]{\thepage}
\fancyfoot[L]{\Date}
\fancyfoot[R]{\Revision}
\fancypagestyle{plain}{
\renewcommand{\headrulewidth}{0pt}
\fancyhf{}
\fancyfoot[C]{\thepage}
\fancyfoot[L]{\Date}
\fancyfoot[R]{\Revision}}
\part[Bose Condensation of Cavity Polaritons and Lasing\\ { - Crossover
Between the Two Regimes}]{Bose Condensation of Cavity Polaritons and
Lasing \\ {\Large Crossover Between the Two Regimes}  }
\chapter{Introduction}
\label{BECintr}
\chset{Introduction}

{\it In this Chapter we introduce basic concepts of Bose-Einstein
condensation and briefly review the main theoretical and experimental
work which has been done up to date in the area of BEC of excitons and
polaritons which is relevant for our work. We also give a short
description of the quantum theory of the laser and a general account of
the relationship between a Bose condensate and a laser.}

\section{Bose-Einstein Condensation}
\label{BEC}

Bose-Einstein condensation is a phase transition in which a
macroscopic number of particles all go into the same quantum state
\cite{bec}. This property place the Bose condensate as a separate state
of matter, distinct from all other states in which the occupation of
any single-particle eigenstate is negligible in comparison to the
total particle number. This macroscopically occupied state has
remarkable properties. In an extended system macroscopic occupation of
one state implies that the wave function of a single particle can
extend over macroscopic distances throughout the system and thus there
is a long-range order in the phase of the condensate. Long-range phase
coherence means that the local phase at every point in the condensate
has a well define value. Since the Hamiltonian does not depend on the
phase of the wavefunctions such phase locking implies broken gauge
symmetry. A well defined phase, according to the Heisenberg uncertainty
principle, causes the number of particles not to have a definite value
and thus the condensate has to be a coherent state and not a number
state.

The only experimental realisations of BEC are liquid Helium and
trapped atomic gases. A superconductor is yet another example of a
macroscopic coherent state. However the fermionic structure of Cooper
pairs in a superconductor affects some of its properties, and thus this
state is sometimes distinguished from the BEC in a strict sense.
It would be of a great fundamental interest to add a few more
experimental realisations to this short list. Liquid Helium and
trapped atomic gases are probably not the most suitable systems for
practical applications and a realisation of a BEC in a solid - state
system would be more practical.

Neutral electronic excitations in a solid can have a bosonic nature, and
thus could be candidates for Bose condensation. In semiconductors
the electron in the conduction band and the hole in the valence band
interact via an attractive Coulomb interaction, and can form a bound state
called exciton. An exciton, as a particle build up from two fermions, is a
boson and thus has the properties which result from Bose statistics.

There are, however, some practical difficulties concerning BEC of
excitons.  In semiconductors the Coulomb interaction between the
electron and the hole within exciton is screened by other carriers and
thus the binding energy of exciton is much weaker than that of the
analogous hydrogen atom. Therefore excitons can only exist at low
temperatures or in low-dimensional structures, where their binding
energy is enhanced.  Another fundamental difficulty is that the large
population of an externally excited excitons is not a real ground
state of semiconductor. Excitons are not conserved particles and can
recombine emitting photons. However, if the lifetime of the exciton is
long in comparison with the thermalisation time a quasi-equilibrium
Bose condensate could be observed. Thus materials with very weak
light-matter coupling would be good candidates for such condensates.
We will give a brief review of experimental advances in this area in
Section \ref{exccond}.

The strong light-matter interaction does not always have to be
disadvantageous for the formation of the condensate. At certain
conditions it is the strong dipole coupling which would favour the
condensation. One of the examples is a coherently excited
semiconductor. In experiments where electrons and holes are created by
an external coherent light the non-equilibrium driven condensate would
form. Coherence in such systems does not appear spontaneously but is
inherited from an external coherent pump. Another example would be to
prevent the coherent excitations from escaping by placing the sample
in an optical cavity. If the interaction between atoms or excitons in
semiconductors and light is strong the new quasiparticles, called
cavity polaritons, would form. Polaritons as a coupled states of
photons and electronic excitations are bosons and thus similarly as
pure excitons could Bose condense. Bose condensation of excitons and
polaritons has been a subject of extensive experimental investigations
in recent years which we will briefly review in Section
\ref{polariton} and Chapter \ref{concl}.

BEC was first described for a non-interacting, ideal gas of free
bosons. This description still remains a standard textbook theory
despite that it does not explain all basic properties of real Bose
condensates stated in the first paragraph of this Section. However,
the instability of the normal state is present even in the
non-interacting Bose-gas and thus this model gives the first insight
into the phenomena of Bose condensation.

For an ideal gas of bosons with mass $m$ and particle number $N$, the
occupation of a single eigenstate with energy $\epsilon_k$ is given by
a Bose distribution
\begin{equation}
n_{\bf k}=\frac{1}{e^{\beta(\epsilon_k-\mu)}-1}, \nonumber
\end{equation}
where $\epsilon_k=\hbar^2 {\bf k}^2/2m$, $\mu \leq 0$ is a chemical
potential and $\beta=1/kT$.
The total number of particles in the system is
\begin{equation}
\sum_{\bf k} n_{\bf k} = N,
\label{eq:sum}
\end{equation}
and thus the density of particles is
\begin{equation}
n= \frac{N}{V} = \frac{1}{(2\pi)^3}\int
\frac{1}{e^{\beta(\epsilon_k-\mu)}-1}d {\bf k}. 
\label{eq:nonint}
\end{equation}
It can be noticed that the integral (\ref{eq:nonint}) cannot account
for all the particles in a system at all $T$ and $n$, since $\mu$ must
have an upper bound of zero for the distribution function to be
defined at all energies. At a given temperature the density of
particles which corresponds to $\mu=0$ is a critical density which can
not be exceeded within the equation (\ref{eq:nonint}). This implies
that for densities higher that this critical density the summation in
equation (\ref{eq:sum}) can not be simply replaced by an integral and
the only way to accommodate all the particles in the system is by
placing the macroscopic number of bosons into a small, non-macroscopic
number of states.

The theory of a non-interacting Bose gas explains the macroscopic
occupation of a small number of quantum states and gives the density
and the temperature of a transition. However, this theory does not
explain why all the excess particles should occupy only one
single-particle state and not a few states close to the ground state
\cite{nozieresbec}. Another problem is that for the non-interacting
boson model the condensate would be a number state of the form
\begin{equation}
(\psi^{\dagger})^N|vac\rangle.
\label{eq:number}
\end{equation}  
The wavefunction (\ref{eq:number}) does not contain the phase
coherence characteristic of real condensates.

Interactions between particles would resolve both problems discussed
in the previous paragraph. Even arbitrarily weak interactions can
create a macroscopic energy difference between different
macroscopically occupied eigenstates, whose energies would be very
close to the ground state in the absence of interactions. Interactions
thus favour the occupation of the single lowest energy eigenstate
only. In the presence of interactions the condensate wavefunction will
no longer be a number state but will become a coherent state of the
form
\begin{equation}
e^{\lambda \psi^{\dagger}}|vac\rangle.
\label{eq:coh}
\end{equation} 
The interaction mixes the single particle states and thus the total
energy of the system can be lowered with respect to the energy in a
number state by taking into account the interference terms between
different single-particle eigenstates within the coherent state
wavefunction. Coherent state has a fixed phase but allows for the
fluctuation of the condensate population. Only the average particle
number in the condensate is well defined and equal to $\lambda$ (eqn
\ref{eq:coh}).

The Bogoliubov theory of the weakly interacting Bose gas is used to
describe atomic condensates. Other condensates mentioned at the
beginning of this section have additional complications which can not
be accounted for by this theory. The interactions in liquid Helium are
far too strong to be described by the weakly interacting gas model. In
the case of superconductors the internal fermionic structure of the
Cooper pairs cannot be neglected. Excitons and polaritons could be
treated as structureless bosons only at very low densities, probably
to low to acquire phase coherence. For higher densities, similarly as
in superconductors, the internal fermionic structure has to be
included.

In the next few sections we will review the existing theories and
experiments concerning quasi-equilibrium exciton condensation,
coherently driven condensates, polariton stimulated scattering and
equilibrium isolated polariton condensation.

\section{Condensation of Excitons} 
\label{exccond}

In materials where the light-matter interaction is very weak, for
example $Cu_2O$ or $CuCl$, the electronic excitations have long
life-time. This time might be long enough to achieve a thermal
equilibrium and to acquire a phase coherence due to the mutual
interactions. Thus we might expect a quasi-equilibrium condensate.

Bosons on an atomic energy scale, apart from photons, are always
composite, build up from an even number of fermions. In the case of
$^4He$ or any other atomic or molecular system it is very difficult
experimentally to compress them to such high densities that the
underlying fermionic degrees of freedom would play an important
role. In the case of electronic excitations, such as for example
excitons in semiconductors, due to their large radii and the
possibilities of creating them by external excitations, the high
densities can be easily realised and the underlying fermionic nature
can play an important role.

The major nonlinearities arising with increasing density are screening
and the saturation of the underlying fermionic states, the so called
``phase-space filling'' effect. At low densities, when the exciton
radius is small compared with the distance between excitons, the
condensate would resemble a BEC of point-like bosons, while in the
opposite limit, where excitons overlap and the fermionic states
saturate, the system would form a BCS-like coherent pair-state, called
an {\it excitonic insulator} in semiconductor physics. In other words
we have a transition from a real-space pairing of tightly bound
excitons to a momentum-space pairing of weakly correlated and
overlapping electron-hole pairs.

By analogy to superconductivity Keldysh, Kozlov and Kopaev developed a
theory \cite{keldysh1,keldysh2} which smoothly connects these two
regimes of densities. Excitons are treated not as structureless bosons
but as electrons and holes interacting via the attractive Coulomb
potential. Introducing $b_{\bf k}$ and $a_{\bf k}$ as annihilation
operators for an electron with momentum {\bf k} in the conduction and
valence bands respectively, the creation operator for an exciton is
\begin{equation}
D^{\dagger}=\sum_{\bf k} \phi_{\bf k} b^{\dagger}_{\bf k}a_{\bf k}.
\nonumber
\end{equation}
$\phi_{\bf k}$ is the internal orbital wavefunction and remains close
to the single exciton form at low densities, where exciton-exciton
interactions do not affect the internal structure of the exciton. As we
discussed in the Section \ref{BEC}, the Bose condensate is a
macroscopically occupied state with long-range phase coherence. Thus
we expect the wavefunction of an exciton condensate to be a coherent
state of the form 
\begin{equation} 
|\Phi_{\lambda}\rangle=e^{\lambda
D^\dagger}|vac\rangle=\prod_{\bf k} e^{\lambda\phi_{\bf k}
b^{\dagger}_{\bf k}a_{\bf k}} |vac\rangle.
\label{eq:exbec}
\end{equation}
Expanding the exponential in the equation (\ref{eq:exbec}) and taking
into account the exclusion principle for fermionic operators
$b^{\dagger}$ and $a$, the condensate wavefunctions would have the form
\begin{equation} 
|\Phi_{\lambda}\rangle=\prod_{\bf k} ( 1 + \lambda \phi_{\bf k}
 b^\dagger_{\bf k} a_{\bf k})|vac\rangle.
\label{eq:exbec2}
\end{equation}
Equation (\ref{eq:exbec2}) is a special case of the BCS wavefunction
for the coherent state of Cooper pairs in the superconductors
\begin{equation}
|u,v\rangle = 
\prod_{\bf k} ( u_{\bf k}  + v_{\bf k}b^{\dagger}_{\bf k} a_{\bf k} )
|vac\rangle,
\label{eq:becform}
\end{equation}
where
\begin{equation}
|u_{\bf k}|^2+|v_{\bf k}|^2=1.
\label{eq:exclusion}
\end{equation}
The wavefunction (\ref{eq:becform}) provides a smooth transition
between Bose-condensed excitons at low densities and BCS-like
collective state of electron and holes (so called excitonic insulator)
at high densities; at very high densities it approaches an uncondensed
electron-hole plasma. This form of the wavefuncion was widely used to
describe BEC of excitons, particularly by Comte and Nozi{\`{e}}res
\cite{nozieresex1}.

\begin{figure}[htbp]
      \begin{center}  
        \leavevmode 
      \epsfxsize=11.0cm 
      \epsfbox{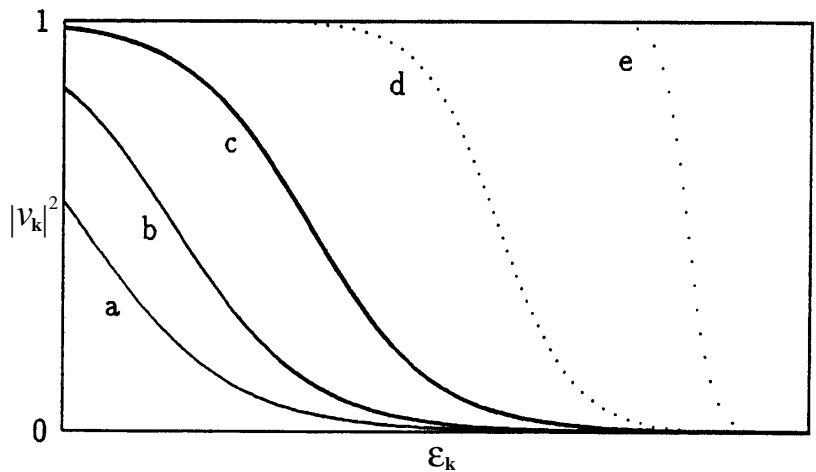} 
        \end{center}
\caption{The ground-state fermion distribution $|v_{\bf k}|$ as a
        function of the electron-hole pair's internal-motion kinetic
        energy $\epsilon_{\bf k}$ for various densities, taken from
        \cite{moscalenko}.}
\label{fig:keldysh}
\end{figure}

The evolution of the state (\ref{eq:becform}) as the density grows can
be seen in Figure \ref{fig:keldysh} . 
The behaviour of $|v_{\bf k}|$ is shown as a function of the energy
$\epsilon_{\bf k}$ for different densities. $\lambda$ is an average
number of particles in the state, fixed by the density, while the
shape of the $v_{\bf k}$ shows the internal structure of the
electron-hole pair. At low densities where $u_{\bf k}\approx 1$ and
$v_{\bf k}\approx \lambda \phi_{\bf k} \ll 1$, $|v_k|$ scales with a
constant shape as a function of density (curves a and b in
Fig. \ref{fig:keldysh}) and describes essentially the single exciton
wavefunction.

With increase in the density the $|v_{\bf k}|$ finally approaches 1
(curve c in Fig. \ref{fig:keldysh}) and due to the exclusion
principle, expressed by the equation (\ref{eq:exclusion}), cannot
increase any further. In order to accumulate more particles it is
necessary to occupy more energetic states in {\bf k}-space (curves d
and e in Fig. \ref{fig:keldysh}). This saturation occurs when excitons
start overlapping, and the resulting ground state becomes BCS-like
collective state, characterised by a small energy gap in the excitation
spectrum which decreases with density. In the regime of very high
densities $v_{\bf k}$ becomes a step function, $|v_{\bf k}|=1$ if
${\bf k}<{\bf k_F}$, $0$ if ${\bf k}>{\bf k}_F$, characteristic for
electron-hole plasma. Thus, the saturation of the fermionic states
together with screening would eventually suppress the condensation at
high densities.

Another enemy of the excitonic condensation are impurities present in
semiconductor. The influence of impurities on the excitonic insulator
was discussed by Zittartz \cite{zittart}. He studied randomly
distributed normal impurities which, in the case of oppositely charged
electrons and holes, have the same effect as magnetic impurities in
superconductors. He applied the Abrikosov and Gor'kov theory
\cite{abrikosov-gorkov} to the excitonic insulator case and showed
that beyond a critical impurity concentration the excitonic phase
cannot exist. Close to the critical concentration there is a gapless
excitonic insulator analogous to a gapless superconductor.

Experimental work on Bose condensation of excitons rely on looking at
systems with a long lifetime for recombination such as $Cu_2O$ and
$CuCl$ or dipole excitons in coupled quantum wells. A very good review
of the recent advances in this area, both the experimental and the
theoretical, can be found in references \cite{moscalenko, bec}.

The evidences for the existence of Bose condensates in $Cu_2O$ are
based on spectral analysis or detecting superfluidity of excitons.
Lin and Wolfe \cite{lin} observed lineshapes in photoemission that were
consistent with strong degeneracy and possibly consistent with BEC of
excitons. Experiments by Mysyrowicz and co-workers
\cite{mys1}-\cite{mys5} reported an anomalous ballistic transport of
excitons over large distances which was argued to be a sign of
superfluidity; these have been criticised as possibly being due to
phonon-wind transport \cite{wind1}-\cite{wind3}, a purely classical
effect. Most recently \cite{wolfe}, Wolfe retracted the earlier
experiments because the absolute measurements of the exciton density
showed they were too low to give BEC.

A coupled quantum wells system consists of two different, adjacent
planes with electrons in one layer and holes in the other. Applied
electric field keep both kinds of carriers in separate two-dimensional
planes. This arrangement reduces the overlap of the wavefunctions of
the electron and the hole and thus increases the lifetime of excitons.
While the first set of experiments claiming to see Bose statistics in
these systems were retracted, the most recent work by Butov {\it et
al.} \cite{butov1}-\cite{butov3} concludes that there is evidence of
stimulated scattering, and therefore densities reaching the quantum
regime; but there is no evidence for BEC.

One problem with an interpretation of the experimental data, and also
with the control of the condensate, comes from the fact that excitons
decay emitting light, and thus the condensate is only a
quasi-equilibrium phenomena, which can survive only up to the time
scale for recombination. One way to go would be to look for materials
where light-matter interaction is very weak and so consequently
excitons have relatively long lifetime. But the strong exciton-photon
interaction does not necessarily preclude the existence of a
condensate if the coherence is maintained in the photons. This could
be realised in two fundamentally different situations.

The first one is the coherently driven system. In the presence of an
external strong, coherent laser field the non-equilibrium condensate
of electronic excitation can form. Instead of spontaneous symmetry
breaking leading to a condensation the exciton system inherit its
coherence from the external coherent photon field.

The second case is a spontaneous condensation of cavity polaritons
achieved by placing the system in an optical cavity, where photons and
thus the coherence is prevented from escaping.
 
\section{Coherently driven systems }
\label{driven}

The properties of the media in the strong coherent electromagnetic
field were first studied by Galitskii, Goreslavskii and Elesin
\cite{gelesin} then broth into the context of exciton condensation by
Elesin and Kopaev \cite{elesincop} and further developed by
Schmitt-Rink, Chemla and Haug \cite{schmitt}.

In the presence of electromagnetic field with very high intensity the
interaction between electronic excitations and light is particularly
strong. It cannot be tackled by a perturbation theory but as a major
interaction in the system must be treated exactly. The Hamiltonian for
the electronic excitations and their interaction with the external
electromagnetic wave $A$ cos $\omega t $ is the following
\begin{equation}
H =\sum_{\bf k} \epsilon_{\bf k}( b^\dagger_{\bf k} b_{\bf k} - 
a^\dagger_{\bf k} a_{\bf k})+
\Delta_k (b^\dagger_{\bf k} a_{\bf k} e^{-i\omega t}+ 
a^\dagger_{\bf k} b_{\bf k} e^{i\omega t} ) + H_{Coul},
\end{equation}
where $\Delta_k=g_kA$ and $g_k$ is a coupling strength between
matter and light. The $H_{Coul}$ is the Coulomb interaction between
carriers. In the absence of the Coulomb interaction it is possible to
diagonalise this Hamiltonian by means of first a unitary
transformation to change to a representation in which the Hamiltonian
does not depend on time and then a canonical transformation to make it
diagonal:
\begin{equation}
H_0 =\sum_{\bf k} E_{\bf k}( \beta^\dagger_{\bf k} \beta_{\bf k} - 
\alpha^\dagger_{\bf k} \alpha_{\bf k}).
\end{equation}
The new quasiparticles $\beta_k$ and $\alpha_k$ are linear
combinations of $a_k$ and $b_k$ with the dispersion law 
\begin{equation}
E_k=\sqrt{(\epsilon_k-\omega)^2+\Delta_k^2}.
\end{equation}
In the presence of a strong electromagnetic field the electronic
levels get mixed and the relevant description is in terms of new
quasiparticles $\beta_k$ and $\alpha_k$, which are coherent
superpositions of the upper and the lower electronic states. Since
the electron is in a coherent superposition of both levels, the
polarisation of the particles is finite. Thus a macroscopic coherent
polarisation of the media appears, inherited from the coherent pump
field.

The new quasiparticles $\beta_k$ and $\alpha_k$ are subject to
other interactions of a different origin (Coulomb interactions,
collisions with phonons or impurities). These additional interactions
can be treated perturbatievly if they are weak in comparison with
electron - photon coupling, and the kinetic equations for the
quasiparticles can be obtained. If the other interactions become
comparable with light-matter coupling, the quasiparticle picture breaks
down and we are back to the usual weak regime of light-matter
interactions.

Schmitt-Rink {\it et al} \cite{schmitt} studied the behaviour of a
coherently driven system for a wide range of pump intensity (i.e the
density of electronic excitations) for two cases with and without the
Coulomb interaction using more rigorous Keldysh non-equilibrium Green
function techniques. They obtained all regimes of real-space and
momentum space pairing discussed in the Section \ref{exccond}. At high
pump intensities the Coulomb interaction gives only a small correction
to the dominant processes induced by the coherent pump filed. In
contrast at small pump intensities the Coulomb interactions play an
important role and leads the electrons and holes to form bound
excitons which are then coherently driven by the field.

Due to the external source of a coherence the coherently driven
condensate is much more robust than the equilibrium excitonic
insulator and thus exist for all temperatures and densities.
Screening and phase-filling effects, which suppress the equilibrium
excitonic condensate at high densities, are also present in the case of
coherently driven systems. However, high densities in this case
correspond to a strong pump field which dominates the physics, and the
gapped condensate is present.

Schmitt-Rink {\it et al} \cite{schmitt} also point out that their
results, in the case without the Coulomb interaction, are equivalent
to those obtained for an ensemble of independent two-level atoms as
the optical transition with different { \bf k} decouple. We will come
back to this point in Section \ref{model} when we introduce the
two-level oscillators model for the electronic excitations.

\section{Polariton Condensation}
\label{polariton}

Another way to prevent the exciton condensate from decaying is to
place the media in the cavity, where photons are confined and thus
prevented from escaping. Similarly as in Section \ref{driven}
strong light-matter interaction mixes the photon and exciton states
and causes the formation of the new quasiparticle cavity-polariton.
Polariton can be seen as a superposition of a plane wave of the
electromagnetic field and a plane wave of polarisation. It was first
introduced by Hopfield \cite{hopfpol} to describe the light in a
bulk material. Cavity polaritons have been experimentally observed for
atoms \cite{atompol}, quantum wells \cite{cavpol} and bulk excitons
\cite{bulkcavpol}, excitons in organic semiconductors
\cite{organicpol1, organicpol2} and exciton complexes \cite{chargedpol}.
 
Quantum wells confined in a planar microcavity is a particularly good
system to observe the coherence effects. The binding energy of ideal
quantum well exciton is four times that of the bulk exciton and the
photon-matter interaction in a microcavity is particularly strong, as
the decoherence effects are reduced, allowing to study the so called
strong-coupling regime.

Microcavity polaritons are subject of intensive experimental research
in recent years. A strong-coupling polariton effect for excitons
coupled to photons was first observed in microcavity GaAs/AlGaAs
quantum well by Weisbuch {\it et al} \cite{cavpol} in 1992. Since then
there were many experiments studying the properties of cavity
polaritons in various systems out of which the most interesting is the
observation of stimulating scattering for polaritons \cite{dang} -
\cite{angle-cw-stimscat}.

Stimulating scattering, similarly as BEC, is an effect of quantum
statistics of indistinguishable Bose particles. The scattering rate
into a particular final state is proportional to a factor $(1+N_f)$,
where $N_f$ is the occupation number of the final state. Thus,
particles already present in the particular eigenstate enhance the
transitions of other particles into this eigenstate. This process for
photons leads to a lasing action while for polaritons it can give an
increase by orders of magnitude of the total scattering rate even in
the normal state. This effect can be observed by measuring the
photoluminescence from an optically excited semiconductors and indeed
in recent experiments \cite{dang} - \cite{angle-cw-stimscat}
authors observe a nonlinear build-up of a lower polariton population
as a pumping intensity is increased.  There is a clear threshold in
the pump intensity where the stimulated scattering starts. Above this
threshold a large occupation of the lower polariton states is
observed. This population is usually far from thermal equilibrium with
few exceptions where the thermalised emission is claimed
\cite{bulkboser}.

Stimulated scattering of polaritons causes macroscopic occupation of a
single quantum state and in this way is similar to Bose condensate.
However BEC is characterised by a phase coherence which is unlikely to
be present in these experiments. The experiments are performed in the
very low density limit and thus a considerably long time is required
for the interactions to produce a phase coherence. This time is most
likely to be longer that the lifetime of the cavity mode. In addition
there are also other pair-breaking decoherence processes, which we
will show in this work to be more destructive at low densities. In
present experiments the attempt to increase the density of polaritons
results not in the formation of the condensate but in a switching into
the weak-coupling regime and lasing.  Thus definite evidence for BEC
of polaritons has not been given.

However, the observation of the stimulating scattering for polaritons
suggests that another effect of bosonic statistic, BEC should be in
principle possible in these systems. The issue is not whether or not
excitons are bound or ionised into a plasma, but whether the
decoherence suppresses the excitonic polarisation. We will show in
this work that indeed the regime of BEC exists for sufficiently large
densities and low decoherence. Thus the problem is to experimentally
achieve the right regime. We will come back to this issue in more
details in the Chapter
\ref{concl}.

Now we will present the existing theory of isolated polaritons in
which the decoherence and dissipations processes are ignored. First we
discuss the low density regime and then the generalised model for
all densities.

\subsection{Low Density Regime}
\label{lowdens}

\begin{figure}[htbp]
      \begin{center}  
        \leavevmode 
      \epsfxsize=9.5cm 
      \epsfbox{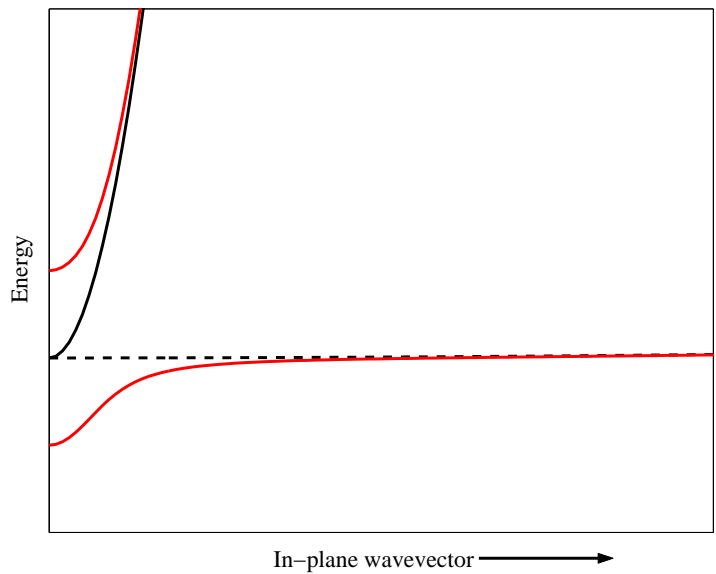} 
        \end{center}
\caption{Illustration of the polariton (red), photon (solid black),
and exciton (dashed black) dispersions for an ideal quantum well
embedded in a planar microcavity, taken from the reference
\cite{paul-phd} }
\label{fig:cavpol}
\end{figure}

The first approximation usually made is to treat the excitons as
structureless bosons and thus the interacting exciton-photon system is
nothing more than two coupled harmonic oscillators. The Hamiltonian for
structureless excitons dipole coupled to photons after the
rotating-wave approximation is 
\begin{equation}
H =\sum_{\bf k} \epsilon_{\bf k} D^\dagger_{{\bf k}} D_{{\bf k}}+
    \omega_{\bf k}  \psi^\dagger_{{\bf k}} \psi_{{\bf k}}+
g_{\bf k}(D^\dagger_{{\bf k}} \psi_{{\bf k}} + \psi^\dagger_{{\bf k}} 
D_{{\bf k}}).
\label{eq:lowdens}
\end{equation}
The Hamiltonian (\ref{eq:lowdens}) can be diagonalised using the
canonical transformation
\begin{displaymath}
\begin{pmatrix} P_{\bf k}^1 \\ P_{\bf k}^2 \end{pmatrix} = 
\begin{pmatrix}\cos \theta_{{\bf k}} & \sin \theta_{{\bf k}} \\
\sin \theta_{{\bf k}} & -\cos \theta_{{\bf k}}
\end{pmatrix}\begin{pmatrix} D_{{\bf k}} \\ \psi_{{\bf k}}
\end{pmatrix}
\end{displaymath}
to the following form
\begin{displaymath} 
H = E^1_{\bf k}P^{1 \dagger}_{\bf k} P^{1}_{\bf k} + 
E^2_{\bf k}P^{2 \dagger}_{\bf k}  P^{2}_{\bf k},
\end{displaymath}
where $P^1_{\bf k}$ and $P^2_{\bf k}$ are the two polariton levels and
are linear combinations of exciton and photon $P^{1,2}_{\bf
k}=u^{1,2}_{\bf k}D_{\bf k}+v^{1,2}_{\bf k}\psi_{\bf k}$ with the
dispersion relation
\begin{equation}
E^{1,2}_{\bf k}=\frac{1}{2}\left[\epsilon_{\bf k}+\omega \pm 
\sqrt{(\epsilon_{\bf k}-\omega)^2+4g_{\bf k}^2}\right]. 
\label{eq:pol-disp}
\end{equation}
The two-dimensional polariton energies are illustrated in
Fig. \ref{fig:cavpol}.
The first estimate in studying the BEC of polaritons is to treat them
as weakly interacting, structureless bosons and to use the theory of
atomic condensates to predict the transition temperature and
density. However one can easily notice that this picture is far from
being complete. There are several effects which are important for the
formation of the condensate and are not included within the above
treatment.

1){ \it Phase space filling effect}. With the increase in the
   excitonic density saturation of the underlying fermionic states
   becomes important. This saturation takes place for densities at
   which excitons start overlapping. Phase space filling effect is not
   included in the model of polaritons described at the beginning of
   this Section.

2){ \it Disorder.} All theories for exciton and polariton condensation
   as well as driven condensates discussed so far consider ideal and
   infinite structures, where bare electronic excitations are
   propagating states with a well defined momentum. However, the real
   structures are far from being perfect and contain a high level of
   disorder caused by fluctuations in thickness of quantum well and
   alloying concentration or presence of impurities. The disorder
   potential is usually weak with respect to the binding energy of
   excitons and thus does not destroy them. However, the exciton centre
   of mass wavefunction would be localised in the minima of the
   disorder potential rather then propagating through the sample. The
   correlation length of the disorder potential and thus the spatial
   extend of the exciton centre of mass wavefunction vary typically
   from a few to about ten exciton Bohr radius \cite{savona}. Thus a
   localised exciton picture might be more relevant for these
   structures.

3){ \it Screening.} The Coulomb interaction between electron and hole
   in the exciton is screened by other excitations. This screening
   increases with the increase in density leading to a complete
   dissociation of excitons at very high densities and thus the
   decrease in the strength of the exciton-photon interaction.

4){\it Decoherence effects.} Apart from the exciton-photon interaction
   there are other processes present in the cavity. Excitons interact
   with phonons and impurities, they are subject to collisions and
   pumping processes. Despite the advances in the fabrication of
   microcavities the mirrors are not perfect and the photons can
   escape. These processes could cause the suppression of the
   condensate and the strong coupling regime altogether.
 
\subsection{Generalisation of Polariton Condensation to Include Phase
Space Filling Effect }
\label{paul}
 
The first two of the effects described at the end of the Section
\ref{lowdens}: the phase space filling effect and the disorder were 
studied by Eastham and Littlewood \cite{paul,paul-phd}. In contrast to
the propagated excitons model they use the generalised Dicke model
widely applied in quantum optics to study the radiative decay of
atomic gases and lasers. The electronic excitations are treated as an
ensemble of two-level oscillators dipole coupled to one mode of the
electromagnetic filed. The upper level corresponds to the excited
state of an atom or the presence of a localised exciton at a given
site in the material, while the lower state corresponds to the ground
state of the atom or the lack of an exciton. The two-level oscillators
interact through the common photon mode and all other direct
interactions are neglected. The Dicke model is generalised to include
the distribution of excitonic energies and coupling strength and the
constrain in the number of excitations is imposed. The Hamiltonian for
this model is
\begin{equation}
H =  \sum_{j=1}^N \epsilon_j (b^\dagger_jb_j-a^\dagger_ja_j)
          + \omega_c \psi^\dagger\psi + 
  \sum_{j=1}^N \frac{g_j}{\sqrt{N}}(b^\dagger_ja_j\psi +\psi^\dagger
          a^\dagger_jb_j),
\label{eq:H0paul}
\end{equation}
where $b$ and $a$ are fermionic annihilation operators for an electron
in an upper and lower states respectively and $\psi$ is a photon
bosonic annihilation operator. The sum here is over the possible sites
where an exciton can be present (different molecules or localised
states in the disorder potential) and not over the momentum like in
the case of propagating excitons. The fermionic structure is included
through the single occupancy constrain for both levels of the
two-level oscillator expressed in the commutation relations for the
$a$ and $b$ fermionic operators. Thus this model includes a phase
space filling type effect and the disorder, and does not include
screening and decoherence effects.

In real structures different types of excitons could be present. The
exciton could be localised in a minimum of the disorder potential,
some traps could be big enough so that a few excitons can be present
on one side, some excitons could extend for large distances and
finally some propagate in the sample. Depending on the material and
the growth condition of the sample certain types would be
dominant. The model discussed in this Section gives very good
description of the tightly bound, Frenkel excitons which can be
treated as a point-like compared with a wavelength of light, localised
by the disorder or bound on impurities or molecular excitons in
organic materials. This model describes the opposite limit to the
ideal structures with propagating exciton centre-of mass wavefunctions
but recent experiments in quantum wells \cite{gammon} - \cite{bonadeo}
suggest that it might be more relevant, at least at low densities.

Eastham and Littlewood consider a thermal equilibrium of this model at
fixed number of excitations (excitation density) in the cavity equal to
$n_{ex}$ ($\rho_{ex}$), where 
\begin{equation}
n_{ex}= \frac{1}{2} \sum_{j=1}^N
(b^\dagger_jb_j-a^\dagger_ja_j) +
\psi^\dagger\psi=N\rho_{ex}.  
\label{eq:nexpaul}
\end{equation}
This corresponds to the situation, where system is initially excited with
an external pump and then isolated so that the excitations can not
decay out of the cavity, and the thermal equilibrium can be achieved.

They use functional path integral techniques and show that the
mean-field solution is well controlled by the number of sites
available, N, with fluctuations of order of 1/N and thus {\it exact} in
the limit where $N \to\infty$. They show that the wavefunction
\begin{equation}
|u,v\rangle = 
\prod_{j} e^{\lambda \psi^{\dagger}}( u_{j}  + v_{j}b^{\dagger}_j a_j ) 
|vac\rangle \nonumber
\end{equation}
is an exact ground state for $N \to \infty$.

\begin{figure}[htbp]
\begin{center}
\includegraphics[width=10cm,clip]{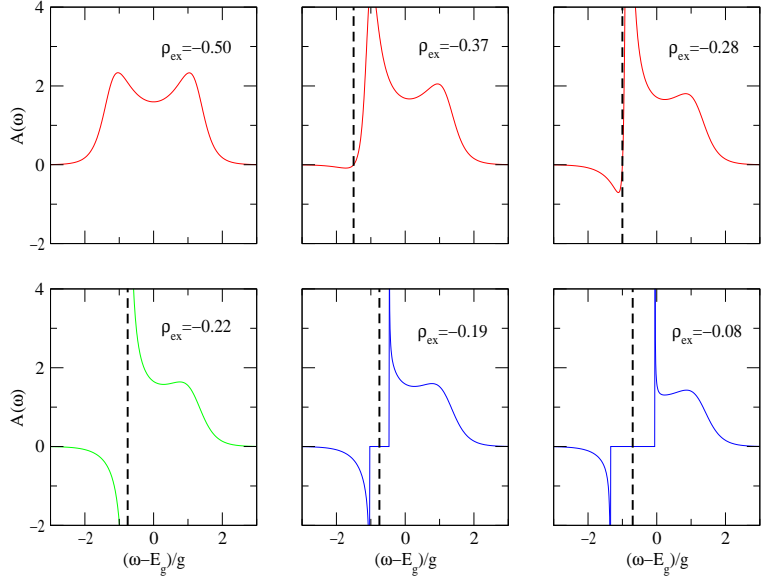}
\end{center}

\caption{The photon spectral functions $A(\omega)$ for polariton
	system (+ve values mean absorption, and -ve emission) at
        different excitation densities $\rho_{ex}$. The top row of
        plots (red curves) are in the normal state, the bottom left
        hand plot (green curve) at the transition and the remaining
        plots (blue curves) in the condensed state, taken from the
        reference \cite{paul-phd}. The dashed lines mark the chemical
        potential.}
\label{fig:paulexc}
\end{figure}

Eastham and Littlewood study the ground state and the excitation
spectrum of this model for a range of temperature, excitation
densities, photon-exciton detuning and exciton energy distribution.
They recover the crossover from the structureless polariton condensate
to the BCS-like coherent state, where saturation effects are
important. In contrast to the excitonic insulator the BEC of
generalised polaritons is present even at very high excitation
densities due to the presence of highly occupied photon mode. At low
densities the condensate is more exciton-like, while at high densities
it has more photon-like character. At zero temperature the system is
always condensed as the decoherence processes are not included in the
model. The condensate has a gap in the excitation spectrum of 
magnitude $4g|\lambda|$, where $\lambda$ is an amplitude of the
coherent photon field and $g$, as usual, is the coupling constant between
photon and exciton (see Figure \ref{fig:paulexc}).

Their work shows that although the phase space filling effect causes
the collapse of the splitting between the two polariton peaks in the
normal state with increasing density, it does not destroy condensation
even at very high excitation densities. Due to this saturation effect
in the fermionic space the condensate is becoming more photon like as
the excitation density is increased, but nevertheless the coherence in
the media and the gap in the excitation spectrum are present (see
Figure \ref{fig:paulexc}).

Polariton condensate as a coherent state of photons and electronic
excitations would be a source of a coherent light. The issue arises
how it is different and how it can be distinguished from the ordinary
laser. Two major conclusions can be derived from their theory.
Firstly, in contrast to lasers the coherent light in polariton
condensate would be present even without population inversion.
Secondly, the polariton condensate, unlike the laser, would have a gap
in the excitation spectrum.

Both processes, the polariton condensate and laser, are fundamentally
different and appear in different regimes of the light-matter
interaction. Polariton condensate is a condensate of strongly coupled
modes of light and electronic excitations. Laser is a weak-coupling
phenomenon and is essentially a coherent state of photons created by
stimulated emission from an inverted electronic population due to the
strong pumping. Both processes can be present in the same microcavity
systems and the essential light-matter interaction would be described
by the same Hamiltonian (\ref{eq:H0paul}). Thus the laser and the polariton
condensate can be viewed as two different regimes in microcavity. In
the Section \ref{exccond} we discussed the crossover between the
regime of BEC and BCS-like states with a change in the density of the
electronic excitations. Now, the crossover between the strong coupling
regime of a polariton condensate and a weak coupling regime of a laser
is caused by a different phenomena. We will show in the Chapter
\ref{abricosov} that the decoherence processes drive this crossover.

The model studied by Eastham and Littlewood is a closed system model
with the dipole interaction between excitons and photons only. However,
in real microcavities other interactions are also present. There are
elastic collisions between excitons, exciton interactions with phonons
and impurities and radiative decay to modes different that the cavity
mode. Excitons are also subject to pumping and the photon field decay
from the cavity. All these processes would cause the decoherence of
the coherent state favoured by the major exciton-photon interaction.
The relative strength of these processes with respect to the
light-matter interaction will determine the regime in the cavity. In a
strong coupling regime all other interactions are very small in
comparison to the dipole interaction, while in the opposite,
weak-coupling regime the other interactions are comparable to the
exciton-photon coupling.

In this work we would like to address the problem of the influence of
the decoherence processes on the polariton condensate. Decoherence can
drive this system through phase transitions. Thus, we would like to
study different regimes as the decoherence is changed for a range of
excitation densities. This will firstly allow us to study the
stability of the polariton condensate to various decoherence processes,
and then to establish a crossover between an ideal polariton
condensate and a laser.

In order to establish the connection between the polariton condensate
and the laser we will first review the basic properties and a quantum
theory of the laser.

\section{Laser}
\label{laser}

Photon laser is a weak-coupling regime phenomenon in which, according to
the Fermi golden rule, the rate of transition is determined by the
squared matrix element of the perturbation between the initial and
final states. In a laser an ensemble of atoms (excitons, electron and
holes) interacts with a photon field confined in the cavity. This
interaction causes transitions between two electronic levels in the
media stimulated by the photon field and thus proportional to the
number of photons in the cavity. Stimulated emission, proportional
also to the number of atoms in an upper state, results in emission of
a photon with the same phase as the photon which causes the transition
and is thus coherent with the cavity mode. Stimulated absorption results
in an absorption of the cavity mode photon and is proportional to the
number of atoms in the lower state. The transition from the upper to
the lower level can also be caused by vacuum fluctuations and is
called spontaneous emission. It causes the emission of the photon with
a random phase and is thus incoherent with the photon mode in the cavity.
Spontaneous emission is small in comparison to the stimulated emission
if many photons are present in the cavity with a ratio 1:n, where n
is the number of photons in the cavity mode. In order to build up a
macroscopic coherent population of photons stimulated emission has to
overcome absorption and thus an inverted ensemble of atoms is
necessary for a laser action. In order to overcome the cavity losses,
spontaneous emission and all other decay processes and to keep the
population of atoms inverted, a sufficiently strong pumping mechanism
has to be applied.

The pumping mechanism and other decoherence processes acting on atoms
are strong, comparable to the interaction with light. Thus lasers
operate in the regime, where the coherent polarisation of the media is
very heavily damped and the atomic coherence is very much
reduced. A coherent photon field, oscillating at bare cavity mode
frequency, is the only order parameter in the system. The above
properties define the regime considered by all laser theories.

The quantum theory of a laser was developed in the sixties using the
quantum Langevin equations, equations of motion for the density matrix
or master equations \cite{sculzub}. We will present the quantum
Langevin equations' formulation developed by Haken
\cite{haken1,haken2} as it can be easily connected to the formulation
of polariton condensate developed by Eastham and Littlewood.

A laser is described as an open system interacting with an environment.
The system consists of an ensemble of two-level oscillators dipole
interacting with a cavity mode. The Hamiltonian for the system is
given by the equation (\ref{eq:H0paul}). All other processes are treated
as an environment which cause decoherence, dissipation and pumping of
the energy. These processes are of a different physical nature,
depending on the material, and their exact details are not important
for the general laser theory. They can be modelled as baths of
harmonic oscillators, interacting with the system in a way that would
give the same effect as the original interactions. These baths are
assumed to be very large in comparison with the system and thus not
changed by the system. The laser Hamiltonian can be written in a
general form as
\begin{equation}
H = H_S+H_{SB}+H_{B},
\label{eq:Hlaser}
\end{equation}
where $H_S$ is a system Hamiltonian given by equation
(\ref{eq:H0paul}). $H_{SB}$ describes the interactions between the system
and its baths
\begin{multline}
H_{SB} = \sum_{k} g_{\kappa}(k)(\psi^\dagger d_{k}  +
d_{k}^\dagger \psi)+ 
 \sum_{j,k}g_{\gamma_{\uparrow}}(j,k)
(b^\dagger_ja_j c_{j,k}^{\alpha \dagger}+c^{\alpha}_{j,k}
a^\dagger_jb_j)+ \\
 \sum_{j,k} g_{\gamma_{\downarrow}}(j,k)
(b^\dagger_ja_j c^{\beta}_{j,k}+c_{j,k}^{\beta \dagger}
a^\dagger_jb_j) 
+\sum_{j,k} g_{\eta}(j,k)(b^\dagger_jb_j
-a^\dagger_ja_j)(c_{j,k}^{\delta \dagger} +c^{\delta}_{j,k}),
\label{eq:HSBlaser}
\end{multline}
while $H_{B}$ is a Hamiltonian for the baths
\begin{multline}
H_{B}  = \sum_{k}\omega_\kappa(k)d_{k}^\dagger d_{k}+
\sum_{j,k}-\omega_{\gamma_{\uparrow}}(j,k)c_{j,k}^{\alpha \dagger}
c_{j,k}^{\alpha}+ \\
 \sum_{j,k}\omega_{\gamma_{\downarrow}}(j,k)c_{j,k}^{\beta
\dagger} c_{j,k}^{\beta}+ 
\sum_{j,k}\omega_{\eta}(j,k)c_{j,k}^{\delta \dagger}
c_{j,k}^{\delta},
\label{eq:HBlaser}
\end{multline}
where $c_{j,k}^{\dagger}$ and $c_{j,k}$ are the bosonic creation and
annihilation operators, respectively, for the baths. The first term in
equation (\ref{eq:HSBlaser}) gives the decay of the photon field from
the cavity. The second term describes incoherent pumping of atoms,
while the third term describes nonradiative recombinations or
spontaneous emission to the modes other than the cavity mode. The
fourth term gives rise to all the decoherence processes which do not
destroy excitations like collisions, phonons or impurity
scattering. The first three terms in equation (\ref{eq:HSBlaser}),
apart from fluctuations, give rise to the flow of an energy through
the system. In order to obtain a dissipation of the energy from the
system the baths need to be kept at a positive temperature whereas to
model the pumping process the bath must be at negative temperature
which correspond to the ensemble of inverted harmonic oscillators.

From the Hamiltonian (\ref{eq:Hlaser}), using some approximations, it is
possible to derive the quantum Langevin equations for the system
operators or density matrix, where the influence of the baths would be
simplified. We will present this procedure for the field operator only,
as the derivation for the atomic operators would follow in a similar
manner.

The Heisenberg equations of motion for the field and its bath annihilation
operators are
\begin{align}
\label{eq:heis-field}
\frac{d\psi}{dt} &=
  -i\omega_c\psi-i\sum_{j}g_ja^\dagger_jb_j-i\sum_{k}g_{\kappa}(k)d_k \\
\frac{dd_k}{dt} &= -i\omega_\kappa(k)d_k - ig_{\kappa}(k)\psi. ,
\label{eq:heis-bath}  
\end{align}
Integrating the equation (\ref{eq:heis-bath}) we obtain the following
equation for the bath operator
\begin{equation}
d_k(t)=d_k(t_0)e^{-i\omega_\kappa(k)(t-t_0)}-ig_\kappa(k)\int_{t_0}^t 
dt^{'}\psi(t')e^{-i\omega_\kappa(k)(t-t')}.
\label{eq:heis-bath2}
\end{equation}
Substituting equation (\ref{eq:heis-bath2}) into equation
(\ref{eq:heis-field}) we obtain the following integro-differential
equation for the field operator
\begin{multline}
\frac{d\psi}{dt}=-i\omega_c\psi-i\sum_{j}g_ja^\dagger_jb_j
-i\sum_{k}d_k(t_0)e^{-i\omega_\kappa(k)(t-t_0)} \\
-\sum_{k}g_\kappa(k)^2\int_{t_0}^t
dt^{'}\psi(t')e^{-i\omega_\kappa(k)(t-t')}. 
\label{eq:heis-field2}
\end{multline}
The third term in equation (\ref{eq:heis-field2}) does not depend
on the field operators and acts as a random fluctuating force. The
forth term is in general quite complicated and connects the behaviour
of the field operator at time $t$ with its past. In laser theory this
term is simplified using the following approximations:
\begin{itemize}
\item There is a smooth dense spectrum of oscillator frequencies.
\item The coupling constant of the system to the bath and the density
      of states for the bath are smooth, slowly varying functions of
      the frequency of the oscillators.
\item The time scales for the bath degrees of freedom are much shorter than
      for the system degrees of freedom, so that there is no memory in
      the system, and the state of the system does not depend on its
      past. It is the so-called Markov approximation.
\end{itemize}
Transforming $\sum_{k} \to \int d \omega_{\kappa}
D_\kappa(\omega_\kappa)$ and using the above approximations, the last
term in equation (\ref{eq:heis-field2}) can be simplified:
\begin{equation}
-\sum_{k}g_\kappa(k)^2\int_{t_0}^t
dt^{'}\psi(t')e^{-i\omega_\kappa(k)(t-t')}=-\kappa \psi(t),
\end{equation}
where $\kappa=\pi g^2_\kappa(0)D_\kappa(0)$ and $D_\kappa$ is a
density of states for the bath. Introducing the following abbreviation
for the fluctuating force
\begin{equation}
-i\sum_{k}d_k(t_0)e^{-i\omega_\kappa(k)(t-t_0)}=F(t), \nonumber
\end{equation}
the final equation for the field operator is
\begin{equation}
\frac{d\psi}{dt}=(-i\omega_c-\kappa)\psi-i\sum_{j}g_ja^\dagger_jb_j+F(t).
\label{eq:heis-field3}
\end{equation}
Following the same procedure it is possible to derive similar
equations of motion for the atomic operators. The equation for the
atomic polarisation is
\begin{equation}
\frac{da^\dagger_jb_j}{dt}=(-i2\epsilon_j-\gamma_{\perp}) a^\dagger_jb_j+
ig_j\psi(b^\dagger_jb_j-a^\dagger_ja_j) +\Gamma_{j-},
\end{equation}
and for the atomic population inversion
\begin{equation}
\frac{d(b^\dagger_jb_j-a^\dagger_ja_j)}{dt} = 
\gamma_{\parallel}(d_0-b^\dagger_jb_j+a^\dagger_ja_j)+
2ig_j(\psi^\dagger a^\dagger_jb_j-b^\dagger_ja_j\psi)+\Gamma_{j,d},
\label{eq:heis-inv}
\end{equation}
where
\begin{align}
\gamma_{\parallel}&=2(\gamma_{\uparrow}+\gamma_{\downarrow}) &
\gamma_{\perp}&=\gamma_{\uparrow}+\gamma_{\downarrow}+\eta &
d_0&=\frac{\gamma_{\uparrow}-\gamma_{\downarrow}}
{\gamma_{\uparrow}+\gamma_{\downarrow}}. \nonumber
\end{align}
and 
\begin{align}
\gamma_{\uparrow}&=\pi g^2_{\gamma_{\uparrow}}(0)D_{\gamma_{\uparrow}}(0) &
\gamma_{\downarrow}&=
\pi g^2_{\gamma_{\downarrow}}(0)D_{\gamma_{\downarrow}}(0) &
\eta &=\pi g^2_{\eta}(0)D_{\eta}(0). \nonumber
\end{align}
$D_{\gamma_{\uparrow}}$, $D_{\gamma_{\downarrow}}$ and $D_{\eta}$
are the densities of states for the appropriate baths.

All the interactions with an environment together with the
distribution of energies $\epsilon_j$ due to disorder, give additional
broadening to the natural linewidth of the system. This broadening can
be {\it homogeneous} or {\it inhomogeneous} depending on its physical
origin.

{\it Inhomogeneous broadening} takes place when the optical transition
energies are split (in the case of degeneracies) or shifted in a
different way for different sites. This broadening has a Gaussian
distribution and can be included into the model by setting the
two-level oscillators energies $\epsilon_j$ to be Gaussian distributed
around some mean value. The physical origin of this broadening
depends on the material used for a laser medium. It can be caused by
the fluctuations in thickness of a quantum well and alloying
concentration or presence of impurities inserted at non-equivalent
lattice sites.

The {\it homogeneous broadening} corresponds to the situation, where
all sites are broadened in the same way. It can be caused by
collisions with phonons or impurities, or lattice vibrations which are
fast in comparison to the laser process. This broadening can only be
modelled by the means of appropriate dynamic baths.

The influence of baths in the Heisenberg equations of motion, after
the approximations described in this section, manifests itself as decay
constants for the photon field and the polarisation independent of time
and frequency, a pump constant for inversion, and the random fluctuating
forces (noise). The mean values of the noise terms are zero. Thus in the
mean-field approximation equations (\ref{eq:heis-field3}) -
(\ref{eq:heis-inv}) would have only appropriate decay constants and
the random forces would disappear. The atomic equations could be
summed to obtain the equation for the total inversion
$D=\sum_{j}(b^\dagger_jb_j-a^\dagger_ja_j)$, and a total polarisation
$P=\sum_{j}a^\dagger_jb_j$. If we then rescale these equations by the
number of atoms in the cavity N, introducing a rescaled photon field
$\psi/\sqrt{N}$, polarisation $P/\sqrt{N}$ and inversion $D/N$ it will
become apparent that the deterministic terms are of order $N^0$, while
the noise terms are all of order $1/\sqrt{N}$. In realistic systems the
number of atoms is very large and thus the noise terms are small
compared with the deterministic terms. In a limit where $N \to \infty$
the noise terms become zero and the mean-field approach is exact.

In order to study a regime below the lasing threshold, where the
macroscopic coherent fields are absent, and also to determine the
linewidth of the laser light, fluctuations above mean-field have to be
included. Most of the other properties of lasers above threshold can be
derived using the mean-field approximation, since the number of atoms
is always very large.

\section{BEC and Lasing}

The properties of a laser can be derived from the Hamiltonian
(\ref{eq:Hlaser}) with large interactions with an environment. If in
the same Hamiltonian we set the coupling constants between the system
and the environment to zero, we obtain Hamiltonian (\ref{eq:H0paul})
of which the ground state was shown by Eastham and Littlewood to be a
polariton condensate. Thus by varying the magnitude of the coupling
constants between the system and the baths in the Hamiltonian
(\ref{eq:Hlaser}) we should be able to study all different phases
driven by the decoherence. This should allow us to check the stability
of the polariton condensation to the interactions with the outside
world at a small coupling strength, and to establish the connection
between polariton condensation and lasers as the decoherence is
increased.

In Chapter \ref{lang} we will show that the Langevin equations 
(\ref{eq:heis-field3}) - (\ref{eq:heis-inv}), widely used in laser  
theories, cannot be used to establish this connection. We
will show that the time(frequency) independent decay constants are
critical in these equations and even at arbitrarily small decoherence
strength lead to completely different solutions from those in the absence 
of the environment. We will give physical insight why the approximate
treatment of the baths, in the manner of equations (\ref{eq:heis-field3})
- (\ref{eq:heis-inv}), gives non-physical answers in some regimes. In
Chapter \ref{abricosov} we use Green's function techniques,
similar to Abrikosov and Gor'kov theory of gapless superconductivity,
to treat the influence of the environment in the self-consistent way
and to study different regimes as the decoherence is changed. Finally,
in Chapter \ref{concl} we summarise the conclusions and give an
account of the work still in progress as well as possible future
developments.

\chapter{BEC and Lasing - Langevin Equations}
\label{lang}
\chset{BEC and Lasing - Langevin Equations}

{\it In this Chapter we consider the Langevin equations derived for
the laser system in Section \ref{laser}. These equations become
the equations of motions for the isolated system studied by Eastham
and Littlewood when the coupling to the environment is zero.
Firstly we show that these equations in a steady state lead to the same
solutions obtained by Eastham and Littlewood using 
path-integral methods. Thus the equation of motion is yet another
method of studying the polariton condensate but more importantly it
brings a connection to the laser equations. We then consider the
Langevin equations with a finite coupling to the environment and show
that the fact the decay constants are independent of time (frequency) is
critical in these equations and even at arbitrarily small decoherence
strength lead to completely different solutions than in the absence of an
environment. This criticality is however non-physical and arises only
due to the approximations used in deriving the Langevin equations.}

\section{Mean-Field Langevin Equations}

The Langevin equations for the laser system (\ref{eq:heis-field3}) -
(\ref{eq:heis-inv}) after the mean-field approximation take the
following form:
\begin{align}
\label{eq:mean-field}
\frac{d\langle\psi\rangle}{dt}& = (-i\omega_c-\kappa)
\langle\psi\rangle - i\sum_{j}g_j\langle p_j \rangle \\ 
\label{eq:mean-pol}
\frac{d\langle p_j \rangle}{dt}& = (-i2\epsilon_j-\gamma_{\perp})
\langle p_j \rangle  + ig_j\langle\psi\rangle \langle d_j\rangle, \\
\frac{d \langle d_j \rangle}{dt}& = \gamma_{\parallel}
(d_0-\langle d_j\rangle)+ 2ig_j(\langle\psi^\dagger\rangle
\langle p_j\rangle - \langle p_j^\dagger\rangle \langle\psi\rangle), 
\label{eq:mean-inv}
\end{align}
where $\psi$, $p_j$ and $d_j$ are the photon field, the polarisation and
the inversion of a two-level oscillator. At zero temperature
$<>$ means the quantum-mechanical average in the ground state of the
system. By definition the coherent state is an eigenstate of a bosonic
annihilation operator and thus the average of this operator in the
coherent state is equal to the corresponding eigenvalue which is 
non-zero for an occupied state. The average of an annihilation operator in
an normal state is however zero. Thus the mean-field equations
(\ref{eq:mean-field}) - (\ref{eq:mean-inv}) describe the behaviour of
the coherent part of the photon field and polarisation. As we
discussed in Section \ref{laser} they are exact in the limit $N
\to \infty$ for an isolated condensate and laser system above the laser
threshold.

Equation (\ref{eq:mean-pol}) can be rearranged to obtain
\begin{equation}
\langle\psi\rangle=\frac{(\frac{d}{dt}+i2\epsilon+
\gamma_{\perp})\langle p_j \rangle}{ig\langle d_j\rangle}. 
\label{eq:lang-field}
\end{equation}
Substituting this expression into equation (\ref{eq:mean-inv}) yields
\begin{equation}
\frac{d(d_j^2+4|p_j|^2)}{dt} = 4\gamma_{\perp}(\frac{\gamma_{\parallel}}
{\gamma_{\perp}}d(d_0-d_j)-2|p_j|^2). 
\label{eq:lang-const}
\end{equation}
This equation is a differential equation which
connects the coherent polarisation and the inversion of the two-level
oscillator. By setting $\gamma_{\perp}$, $\gamma_{\parallel}$ and
$\kappa$ to zero in the equations (\ref{eq:mean-field}) -
(\ref{eq:mean-inv}) we can obtain the Heisenberg equations of motion
for the isolated system studied by Eastham and Littlewood. Let us first
consider the isolated case.
\section{Isolated System}
The equation (\ref{eq:lang-const}) for the isolated system becomes
\begin{alignat}{3}
\frac{d(d_j^2+4|p_j|^2)}{dt} &= 0 \qquad & 
&\Longrightarrow \qquad & d_j^2+4|p_j|^2 & = Cons. 
\label{eq:lang-const2}
\end{alignat}
The polarisation of the medium is connected with the transitions
between levels and is equal to the off-diagonal terms of the system
density matrix or the Green function. For the coherent polarisation to
have a non-zero value, the system must be in a mixture of lower and upper
states. When the inversion is exactly 1 (or -1), all atoms are in the
upper (or lower state) and thus the coherent polarisation must
disappear. This condition would be satisfied by constant in 
equation (\ref{eq:lang-const2}) equalling unity and thus the relation
between the coherent polarisation and inversion would become
\begin{equation}
|p_j|^2 = -\frac{1}{4}(d_j-1)(d_j+1).
\label{eq:lang-p_d}
\end{equation}
The dependence of the coherent polarisation on the inversion is
shown in Figure \ref{fig:pol_inv} (blue curve). In the
isolated condensate, unlike in lasers, the coherent polarisation is
present even without the population inversion in the system (even for
$d<0$). We are interested in the steady state solutions to these
equations in which the coherent fields oscillate at the constant
frequency:
\begin{align}
\langle\psi(t)\rangle&=\langle\psi\rangle e^{-\mu t} &
\langle p(t) \rangle&=\langle p \rangle e^{-\mu t}. 
\label{eq:ansaths}
\end{align}
Substituting both the Ansatz (\ref{eq:ansaths}) and equation
(\ref{eq:lang-p_d}) into equations (\ref{eq:mean-field}) and
(\ref{eq:mean-pol}) we obtain
\begin{align}
\label{eq:lang-ans1}
\langle\psi\rangle&=\frac{-1}{\omega_c-\mu}\sum_jg_jp_j \\
\label{eq:lang-ans2}
\langle p_j \rangle&=-sign(\epsilon_j-\mu)\frac{g_j \langle\psi\rangle }
{2\sqrt{g_j^2|\langle\psi\rangle|^2+(\epsilon_j-\mu)^2}} \\
\label{eq:lang-ans3}
\langle d_j\rangle&=\frac{-|\epsilon_j-\mu|}
{\sqrt{g_j^2|\langle\psi\rangle|^2+(\epsilon_j-\mu)^2}}.
\end{align}
From equations (\ref{eq:lang-ans1}) and
(\ref{eq:lang-ans2}) after rescaling the coherent field by
$\sqrt{N}$ ($\langle\psi\rangle\to \langle\psi\rangle \sqrt{N}$) we
obtain the following equation for the coherent field
\begin{equation}
\label{eq:BCSgap}
(\omega_c-\mu)\langle\psi\rangle=\frac{\langle\psi\rangle}{2N}\sum_j
\frac{g_j^2}{E_j}
\end{equation}
where we define
\begin{equation}
E_j=sign(\epsilon_j-\mu)\sqrt{g_j^2|\langle\psi\rangle|^2+
(\epsilon_j-\mu)^2}. \nonumber
\end{equation}
Equations (\ref{eq:lang-ans1}) - (\ref{eq:BCSgap}) are exactly the same
as the zero temperature equations obtained by Eastham and Littlewood
\cite{paul,paul-phd} using path-integral methods and describe the
polariton condensate for a wide range of densities. The structure of
the equation (\ref{eq:BCSgap}) is similar to the BCS gap equation.
\section{System with Decoherence}
In this Section we will examine the case when interactions with the
environment are present. We are again interested in the steady state
solutions of the equation (\ref{eq:lang-const}). It can be shown that
the only steady state solution to this equation occurs when
\begin{alignat}{3}
\frac{\gamma_{\parallel}}
{\gamma_{\perp}}d_j(d_0-d_j)-2|p_j|^2 &= 0 \qquad & 
&\Longrightarrow \qquad & |p_j|^2=\frac{\gamma_{\parallel}}
{2\gamma_{\perp}}d_j(d_0-d_j).
\label{eq:decoh-p_d}
\end{alignat}
In the most ideal case when we have neither spontaneous emission nor
any other decoherence process in the system and the only interactions
with the environment are pumping and the decay of the cavity field we
obtain
\begin{equation}
|p_j|^2=\frac{1}{2}d_j(1-d_j).
\label{eq:decoh-p_d2}
\end{equation}

\begin{figure}[hbtp]
	\begin{center}
	\includegraphics[width=16.2cm,clip]{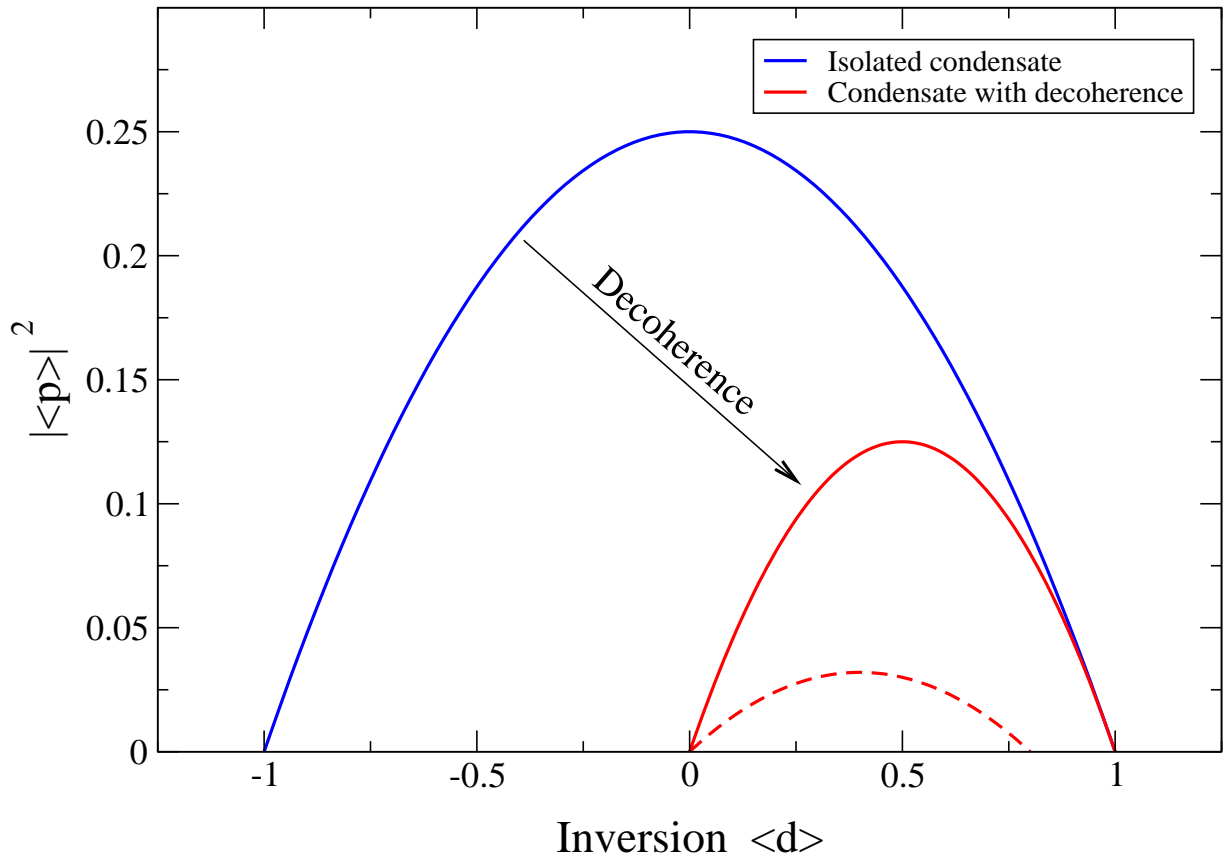} 
	\end{center}
	\caption{Modulus squared of the coherent polarisation as a
	function of an inversion for an isolated condensate (blue) and
	a condensate with an arbitrarily small decoherence (red). The
	solid red
	curve shows the case where only the pumping of excitons and
	the decay of the cavity field are present and is the same for
	all values of the pumping. The dashed curve shows the case
	where all other decoherence processes are present and
	corresponds to the case where $d_0=0.8$ and
	$\gamma_{\parallel}/\gamma_{\perp}=0.4$. Usually in lasers
	$\gamma_{\perp} \gg \gamma_{\parallel}$ and thus the coherent
	polarisation is very small for all values of the inversion.  }
\label{fig:pol_inv}
\end{figure}

We can see from equations (\ref{eq:decoh-p_d}) and
(\ref{eq:decoh-p_d2}) that for an arbitrarily small strength of
the interaction with the environment we obtain completely different
solutions to that given by equation (\ref{eq:lang-p_d}) for an
isolated system. It can be shown that the solutions for an isolated
condensate would decay out when interactions with the outside
world are present. Even for an arbitrary small interaction with the
environment in order to obtain a non-zero coherent polarisation, the
inversion $d$ must be grater than zero which means the system must be
inverted. Figure \ref{fig:pol_inv} shows the comparison between
the isolated condensate (blue curve) and one interacting with an
environment (red curves). The solid red curve shows the most ideal
case when the only interactions are pumping and the decay of the cavity
field. Even in this situation for an arbitrarily small pumping the
system goes from the blue to the solid red curve. The dashed red curve
shows the example when other decoherence processes are present for the
case where $d_0=0.8$ and $\gamma_{\parallel}/\gamma_{\perp}=0.4$. This
is still a very small decoherence. In a usual operating regime for a laser
$\gamma_{\perp} \gg \gamma_{\parallel}$ and thus the coherent
polarisation would be very small for all values of the inversion.

The frequency at which the coherent fields oscillate for an isolated
condensate can be obtained from the density equation
(\ref{eq:nexpaul}). In general this frequency has a very complicated
form but at low densities approaches the polariton energy given by 
equation (\ref{eq:pol-disp}) and is dependent on the coupling strength
between matter and light, $g$ for all densities. In the case where the
interactions with the environment are included within the Langevin
equations (\ref{eq:heis-field3}) - (\ref{eq:heis-inv}) this frequency
can be calculated by comparing the real and the imaginary parts of 
equations (\ref{eq:mean-field}) and (\ref{eq:mean-pol}). In the case
of uniform energies $\epsilon_j=\epsilon$ and coupling constant
$g_j=g$ it takes a simple form
\begin{equation}
\mu=\frac{\gamma_{\perp}\omega_c+\kappa2\epsilon}{\gamma_{\perp}+\kappa},
\nonumber  
\end{equation}
which in the case of the usual laser operating condition $\gamma_{\perp}
\gg \kappa$ leads to $\mu \to \omega_c$. Again it can be seen that
even for an arbitrarily small coupling to the environment this
frequency is completely different to the one obtained for an
isolated condensate. Most importantly it does not depend on $g$. Thus
we can conclude that within the Langevin equation approach with 
constant decay rates even for arbitrary small coupling to the
environment the system goes to the weak light-matter coupling
regime. This surely cannot be physical.

\section{Conclusions}

We have shown that the Langevin equations with constant decay rates
even for an arbitrary small decoherence strength lead to completely
different, essentially laser-type, solutions than for the isolated
system.

In this work we would like to stress that this criticality is not a
physical one and results purely from the approximations applied in
deriving the Langevin equations for this system. This conclusion seems
not to be widely appreciated and even in a relatively recent publication
\cite{wrong} similar decay constants in the equations of
motion for the coherently driven excitonic insulator have been used,
leading to the conclusion that the excitonic insulator phase cannot exist
for an arbitrary small decoherence.

Let us get some physical insight into this problem. In order to obtain
the Langevin equations (\ref{eq:heis-field3}) - (\ref{eq:heis-inv})
all the baths were averaged out before the essential dipole
interaction was taken into account. The ideal condensate has a gap in
the density of states which would still be present for small
decoherence. It is clear that the coherent fields in this regime cannot 
be damped just by constant decay rates independent of frequency as
there are no available states to decay to. As the decoherence is
increased this gap gets smaller and finally is completely suppressed
causing the coherent fields to be strongly damped. In order to
describe this behaviour the decoherence processes have to be treated
self-consistently. The baths have to be averaged out in such a way that
the essential interactions and thus the gap in the density of states is
taken into account.

The equations of motion with decay constants independent of time
and frequency, although perfectly correct in describing the laser where
the decoherence processes are very large, cannot be used to study the
regime with a small decoherence when the gap in the density of states is
present nor to establish a connection between an isolated condensate and
a laser. 

In the next chapter we present an alternative method which takes all
this essential physics into account. By analogy with superconductivity
we use Green function methods similar to the Abrikosov and Gor'kov
theory of gapless superconductivity and treat the decoherence in a
self-consistent way which allow us to move smoothly between an
isolated condensate and other regimes driven by the decoherence.

\chapter{Self-consistent Green's-Function Approach}
\chset{Self-consistent Green's-Functions Approach}
\label{abricosov}

{\it We study the behaviour of a system which consists of a photon
mode dipole coupled to a medium of two-level oscillators in a
microcavity in the presence of the decoherence. We consider two types
of decoherence processes which are analogous to magnetic and
non-magnetic impurities in superconductors. We study different phases
of this system as the decoherence strength and the excitation density
is changed. For a low decoherence we obtain a polariton condensate
with comparable excitonic and photonic parts at low densities and a
BCS-like state with bigger photon component due to the fermionic phase
space filling effect at high densities. In both cases there is a large
gap in the density of states. As the decoherence is increased the gap
is broadened and suppressed, resulting in a gapless condensate and
finally a suppression of the coherence in a low density regime and a
laser at high density limit. A crossover between these regimes is
studied in a self-consistent way similar to the Abrikosov and Gor'kov
theory of gapless superconductivity \cite{abrikosov-gorkov}.}

\section{Model}
\label{model}

\begin{figure}[htbp]
	\begin{center}
	\leavevmode
		\epsfxsize=16.2cm
		\epsfbox{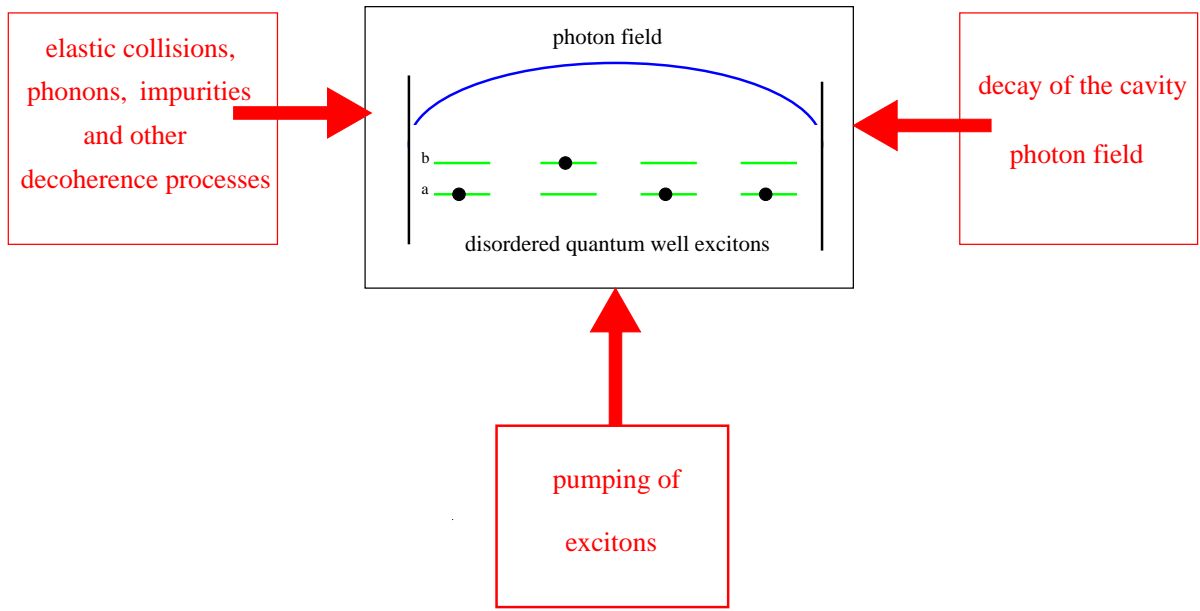}
	\end{center}
	\caption{Sketch of the model studied in this work: the system
	of two-level oscillators dipole coupled to a single cavity
	mode (black box) interacting with various types of environment
	(red boxes).}
\label{fig:model}
\end{figure}

The model we consider in this work is schematically shown in 
Figure \ref{fig:model}.
It consists of a system of two-level oscillators dipole coupled to one
confined cavity mode as described in the Section \ref{paul}. This
system is then subject to various decoherence, pumping and damping
processes described, as in laser theory (Section \ref{laser}), as
baths of harmonic oscillators coupled to the system in a way that
gives the same effect as the original interactions.  

This model includes:
\begin{itemize}
\item major Coulomb interaction between the electron and hole within
      exciton,   
\item phase space filling effect,
\item disorder in the material (inhomogeneous broadening of excitonic
      energies),
\item various types of decoherence effects,
\end{itemize}
and does not include:
\begin{itemize}
\item  screening,
\item  Coulomb interactions between excitons.
\end{itemize}
Thus this model:
\begin{itemize}
\item gives a very good description of tightly bound, Frenkel - type
of excitons localised by disorder or bound on impurities, molecular
excitons in organic materials or atoms in the solid state, 
\item gives a qualitative description within a mean-field approximation
for other types of excitons like Wannier excitons or excitons
propagating in a sample.
\end{itemize}

The second application follows from the fact that the dipole
interaction between excitons and photons is a dominant interaction at
high excitation densities (large photon fields). In the case of a
driven excitonic system Schmitt-Rink {\it at al} \cite{schmitt}
pointed out that for a very large pumping, and thus high excitation
densities, the Coulomb interaction is just a very minor correction to
the dominant dipole coupling. They also state that in the absence of
the Coulomb interaction their results for propagating electrons and
holes are equivalent to those obtained for an ensemble of independent
two-level oscillators as optical transitions with different {\bf k}
decouple. In the case of low densities the dominant Coulomb
interaction, that between an electron and hole within the same
exciton, is taken into account in our model. All other Coulomb
interactions are much weaker and in any case are found within
the mean-field techniques of Keldysh \cite{keldysh1} - \cite{nozieresex1}
for exciton condensation to give rise to a formation of a coherent
excitonic insulator phase and thus would only enhance the dipole
coupling. Therefore the localised two-level oscillator model can be
applied to obtain a qualitative description of the physical behaviour
even for the propagated or weekly bound excitons. It would give
similar predictions to the model based on propagating electrons and
holes with Coulomb interactions treated within mean-field approximations. It
does not include screening and other non mean-fields effects.

We would like to stress that in this work we intend to study a very
general model of the media and consider in details the influence of
various types of decoherence processes in microcavity and different
phases induced by the decoherence and the excitation density. At this
stage we do not want to model any particular media with its complex
interactions which would only make the general picture less clear.
The specific details of the particular media could however be easily
included into our model, we will discuss this possibility in the
Sections \ref{future-decoh} and \ref{future-gen}.

We consider the following Hamiltonian:
\begin{equation}
H = H_0+H_{SB}+H_{B}
\label{eq:H}
\end{equation}
where
\begin{equation}
H_0  = \sum_{j=1}^N \epsilon_j (b^\dagger_jb_j-a^\dagger_ja_j)
          + \omega_c \psi^\dagger\psi + 
        \sum_{j=1}^N \frac{g_j}{\sqrt{N}}(b^\dagger_ja_j\psi +
	\psi^\dagger a^\dagger_jb_j).
\label{eq:H0}
\end{equation}
$H_0$ describes an ensemble of two-level oscillators dipole coupled to
one cavity mode and is exactly the same as the Hamiltonian
(\ref{eq:H0paul}) studied for an isolated condensate by Eastham and
Littlewood \cite{paul,paul-phd}. For the detailed description of this
Hamiltonian see Section \ref{paul}. $H_{SB}$, similarly as in the laser
theory described in the Section \ref{laser} (equation
(\ref{eq:HSBlaser})), contains all interactions with an
environment. In our work we include a more general coupling to the
environment than in the laser theory, allowing the interactions with an
upper level to be different that with a lower level (the fourth and
the fifth terms in the equation (\ref{eq:HSB}) instead of the fourth
term in the equation (\ref{eq:HSBlaser})):
\begin{multline}
H_{SB}=\sum_{j,k} g_{\kappa}(k)(\psi^\dagger d_{k}  +
d_{k}^\dagger \psi)+ \\
 \sum_{j,k}g_{\gamma_{\uparrow}}(j,k)
(b^\dagger_ja_j c_{j,k}^{\alpha\dagger}+c_{j,k}^{\alpha}
a^\dagger_jb_j)+
\sum_{j,k} g_{\gamma_{\downarrow}}(j,k)
(b^\dagger_ja_j c_{j,k}^{\beta}+c_{j,k}^{\beta \dagger}
a^\dagger_jb_j) \\
+ \sum_{j,k} g_{\gamma_{b}}(j,k)b^\dagger_jb_j
(c_{j,k}^{b\dagger} +c_{j,k}^{b})-
\sum_{j,k} g_{\gamma_{a}}(j,k)a^\dagger_ja_j
(c_{j,k}^{a \dagger} +c_{j,k}^{a}). 
\label{eq:HSB}
\end{multline}
The total number of excitations
\begin{equation}
n_{ex}= \frac{1}{2} \sum_{j=1}^N
(b^\dagger_jb_j-a^\dagger_ja_j) + \psi^\dagger\psi,
\label{eq:nex}
\end{equation}
which is the sum of photons and excited two-level oscillators, is
conserved in a steady state. The $H_{B}$ is a Hamiltonian for the
baths as in the equation (\ref{eq:HBlaser}). In the Hamiltonian
(\ref{eq:H}) we have not included terms that couple different sites
(i.e. $a_jb^\dagger_k$ with $k \ne j$) as their influence would be
small in comparison to the coupling within the same site.

The second term in equation (\ref{eq:HSB}) describes an incoherent
pumping of two-level oscillators. The third term contains all the
processes which cause the transition from the upper to the lower level
and thus destroy the electronic excitations such as the decay to
photon modes different to the cavity mode. Finally, the fourth and the
fifth terms describe all the decoherence processes which act on upper
and lower levels respectively but do not change the total number of
excitations in the cavity, for example collisions and interactions
with phonons and impurities. Thus these four terms contain all the
essential groups of decoherence processes in the media. The first term
gives rise to the decay of the photon field from the cavity.

The fourth and the fifth term in equation (\ref{eq:HSB}) can be
expressed as
\begin{equation}
\sum_{j,k} g_{\gamma_1}(j,k)(b^\dagger_jb_j
+a^\dagger_ja_j)(c_{j,k}^{\zeta\dagger} +c_{j,k}^\zeta)+ 
\sum_{j,k} g_{\gamma_2}(j,k)(b^\dagger_jb_j
-a^\dagger_ja_j)(c_{j,k}^{\theta\dagger} +c_{j,k}^\theta).
\label{eq:Hmag}
\end{equation}
The first term in expression (\ref{eq:Hmag}) with the second and the
third terms in the equation (\ref{eq:HSB}) describe the part of a
total interaction which acts on the upper and lower levels with
opposite sign. It can be shown that these processes are analogous to
the pair-breaking, magnetic impurities in superconductors.  We will
refer to them later as the type 1 decoherence processes. The exact
origin of these interactions depends on the details of the
media. Physically these processes correspond to potentials which vary
rapidly in space (on the lengthscale of excitons) or in time. They can be
high energy phonons, collisions or impurities. In the case of
electrons and holes even the normal impurities could be pair-breaking
processes as the electric charge of the impurity acts with the
opposite sign on the electron and the hole \cite{zittart}.

The second term in expression (\ref{eq:Hmag}) contains
interactions which are the same for both levels and are analogous to
normal, non-magnetic, impurities in superconductors. We will call
them type 2 processes. This processes correspond to potentials
which vary very slowly in space in comparison to the size of excitons or
in time. They do not have any pair-breaking effects and lead only to a
broadening of energies.

Since the upper and the lower level are different and contain fermions
with an opposite charge both kinds of interactions would always be
present. In addition to these two types of processes the energies of
the two-level oscillators, $\epsilon_j$, can be inhomogeneously broadened
due to fluctuations in the thickness of quantum well and alloying
concentration or presence of impurities inserted at non-equivalent
lattice sites.
 
The second and the third terms in equation (\ref{eq:HSB}), describing
the incoherent pumping and dumping respectively can, in addition to
having the pair-breaking effect, change the number of electronic
excitations in the cavity and together with the decay of the cavity
photon mode cause the flow of energy through the system. In the steady
state, despite this energy flow, the total number of excitations is
constant in time and proportional to the ratio between the pumping and
the damping processes while the absolute magnitude of all the
processes gives the decoherence strength. For very big pumping and
photon decay rates the system would be out of equilibrium.

In this work we present only the influence of the decoherence on the
system. The implications of a non-equilibrium system is work in
progress. The equilibrium assumption can be completely justified if
the system is pumped slowly, with a rate slower than the
thermalisation rate. This does not put any restrictions on the
strength of the decoherence nor the density of excitations. The
pair-breaking processes described by the first term in expression
(\ref{eq:Hmag}) give exactly the same decoherence effects as the
pumping and damping of two-level systems  preserving 
equilibrium for any strength of the interaction.  The ratio
between the pumping and all the damping processes which can be
arbitrary big even for small absolute values of both will be expressed
in terms of the excitation density $\rho_{ex}$. For rapidly pumped
systems, where the thermal distribution of quasiparticles cannot be
assumed, we study only the influence of the decoherence. 
Non-equilibrium techniques would have to be applied to describe all
the physics. We will address this problem later.

The analogy between superconductivity and an excitonic insulator was
noticed and used in a pioneering work of Keldysh
\cite{keldysh1,keldysh2} and explored later on by others. This analogy
was also stressed by Eastham and Littlewood \cite{paul,paul-phd} in
their work on polariton condensation. The analogy to
superconductivity appears to be useful also in our study of the
influence of the decoherence on the polariton condensate. We can
easily show that the type 1 decoherence processes acting on the
microcavity condensate are analogous to the magnetic impurities in
superconductors while the type 2 processes to the non-magnetic
ones. Thus we can use similar methods to that used by Abrikosov and
Gor'kov in their theory of gapless superconductivity
\cite{abrikosov-gorkov}.

\section{Green's-Function Formulation}
\label{method}

We use the self-consistent Green's function techniques similar to the
Abrikosov and Gor'kov theory of gapless superconductivity
\cite{abrikosov-gorkov,skalski,green}. Introducing the Nambu notation
\begin{displaymath}
\eta_{j}=\left( \begin{array}{c}
b_{j} \\ a_{j}
\end{array}
\right), 
\end{displaymath} 
all the Green functions are 2 x 2 matrices
\begin{displaymath}
G=\left( \begin{array}{cc}
G_{bb} & G_{ba} \\ G_{aa} & G_{ab}
\end{array}
\right) \end{displaymath}
as in the BCS theory. $G_{bb}(j,t-t^{'})=-\langle Tb_j(t)
b^{\dagger}_j(t^{'})\rangle$ and $G_{aa}(j)=-\langle
Ta_j(t)a^{\dagger}_j(t^{'})\rangle$ are the normal, diagonal Green's
functions which give the population distribution of the two-level
oscillators while $G_{ab}(j,t-t^{'})=-\langle
Tb_j(t)a^{\dagger}(t^{'})\rangle$ and $G_{ba}(j,t-t^{'})=-\langle
Ta_j(t)b^{\dagger}(t^{'}) \rangle$ are the anomalous, off-diagonal,
ones which give the coherent polarisation of the media and are
non-zero only in a coherent state.

Since the overall phase of a coherent state is arbitrary, we can
choose the $\langle \psi \rangle$ to be real and thus the Fourier
transform of the zero order Green's function, which is the Green's
function for the Hamiltonian $H_0$ (equation (\ref{eq:H0})) is
\begin{equation}
G^{-1}_0(j,i\omega_n) = i\omega_n-(\epsilon_j-\mu)\tau_3 +
g\langle \psi \rangle \tau_1, \nonumber
\end{equation}
where $\tau_{1,3}$ are the usual Pauli spin matrices. We can introduce
the self-energy $\Sigma$ through the Dyson equation
\begin{equation}
G^{-1}(j,i\omega_n)  =  G_0^{-1}(j,i\omega_n)-\Sigma(j,i\omega_n), 
\label{eq:G1}
\end{equation}
where $G$ is a Green's function for the whole Hamiltonian $H$,
equation (\ref{eq:H}). Now we can write the total Green's function $G$
in the same form as the zero-order Green's function $G_0$
\begin{equation}
G^{-1}(j,i\omega_n) = i\tilde{\omega_n}-(\tilde{\epsilon_j}-\mu)\tau_3 +
g\tilde{\langle \psi \rangle} \tau_1,
\label{eq:G2}
\end{equation}
using the frequency dependent, renormalised $\tilde{\omega_n}$,
$\tilde{\epsilon_j}$ and $\tilde{\langle \psi \rangle}$.  

In this work we consider the decoherence processes given by the
expression (\ref{eq:Hmag}). The self-energy for the type 1 processes
(the first term in the expression (\ref{eq:Hmag})) in a Born
approximation is
\begin{equation}
\label{eq:selfener1}
\Sigma_1(j,i\omega_n) = 
\langle \sum_{k}g_{\gamma_1}(k)(c_{j,k}+c^{\dagger}_{j,k})G(j,i\omega_n)
\sum_{k^{'}}g_{\gamma_1}(k)(c_{j,k^{'}}+c^{\dagger}_{j,k^{'}})\rangle,
\end{equation}
while for the type 2 processes (the second term in the expression 
(\ref{eq:Hmag})) is 
\begin{equation}
\label{eq:selfener2}
\Sigma_2(j,i\omega_n) = 
\langle \sum_{k}g_{\gamma_2}(k)(c_{j,k}+c^{\dagger}_{j,k})\tau_1 G(j,i\omega_n)
\tau_1\sum_{k^{'}}g_{\gamma_1}(k)(c_{j,k^{'}}+c^{\dagger}_{j,k^{'}})\rangle. 
\end{equation}
Using the Markov approximation and assuming that all coupling
constants of the system to the baths, as well as the baths density of
states, are very broad slowly varying functions of frequency we
obtain
\begin{equation}
\Sigma_1(j,i\omega_n) = \gamma_1G(j,i\omega_n),
\label{eq:self1}
\end{equation}
and
\begin{equation}
\Sigma_2(j,i\omega_n) = \gamma_2\tau_1G(j,i\omega_n)\tau_1,
\label{eq:self2}
\end{equation}
where $\gamma_1=g_{\gamma_1}(0)^2N_1(0)$ and
$\gamma_2=g_{\gamma_2}(0)^2N_2(0)$. $N_{1,2}$ are densities of
states for the respective baths.

Substituting equations (\ref{eq:self1}) or (\ref{eq:self2}) into
equation (\ref{eq:G1}) and then comparing with the equation
(\ref{eq:G2}) we can obtain for both cases three equations determining
the renormalised frequency, energy and coherent photon field
\begin{align}
\label{eq:renorm-omega}
\tilde{ \omega_n } &=\omega_n 
+ \gamma_{1,2} \frac{-g\tilde{ \omega_n }}
{\tilde{\omega_n}^2+(\tilde{\epsilon_j}-\mu)^2 + g^2\tilde{\langle \psi
\rangle}^2},  \\
\label{eq:renorm-epsilon}
\tilde{ \epsilon_j } &= \epsilon_j 
- \gamma_{1,2} \frac{-g\tilde{ \epsilon }}
{\tilde{\omega_n}^2+(\tilde{\epsilon_j}-\mu)^2 + g^2\tilde{\langle \psi
\rangle}^2},  \\
\tilde{\langle \psi \rangle} &=\langle \psi \rangle
\mp \gamma_{1,2} \frac{-g\tilde{\langle \psi \rangle}}
{\tilde{\omega_n}^2+(\tilde{\epsilon_j}-\mu)^2 + g^2\tilde{\langle \psi
\rangle}^2}.
\label{eq:renorm-field}
\end{align}
We then have a self-consistency equation for the off-diagonal part of
the Green's function $G$ which is equal to the average coherent
polarisation of the media
\begin{equation}
\frac{1}{N}\langle \sum_j a^{\dagger}_jb_j \rangle = \langle P \rangle = 
\beta^{-1}\sum_{\omega_n,j}
\frac{-g\tilde{\langle \psi \rangle}}
{\tilde{\omega_n}^2+(\tilde{\epsilon_j}-\mu)^2 + g^2\tilde{\langle \psi
\rangle}^2}.
\label{eq:selfcons}
\end{equation}
Both order parameters, the coherent polarisation and the coherent
photon field, are connected through equation
\begin{equation}
\langle \psi \rangle= \frac{-g}{\omega_c-\mu} \langle
P \rangle.
\label{eq:field_pol}
\end{equation}
From equation (\ref{eq:field_pol}) we can see that the ratio
between the two order parameters is essentially determined by the
chemical potential, which in the steady state can be calculated from
the equation (\ref{eq:nex}). A number of electronic excitations
refered to later as inversion can be obtained from the diagonal elements
of the Green's function and is equal to
\begin{equation}
\frac{1}{2}\langle \sum_j(b^{\dagger}_jb_j-a^{\dagger}_ja_j) \rangle =
\beta^{-1}\sum_{\omega_n,j} \frac{-2(\tilde{\epsilon}_j-\mu)}
{\tilde{\omega_n}^2+(\tilde{\epsilon_j}-\mu)^2 +
g^2\tilde{\langle \psi \rangle}^2}.
\label{eq:inv}
\end{equation}

Using the equations (\ref{eq:renorm-omega}) - (\ref{eq:renorm-field})
we can determine the renormalised $\tilde{\omega_n}$,
$\tilde{\epsilon_j}$ and $\tilde{\langle \psi \rangle}$ as a functions
of bare $\omega_n$, $\epsilon_j$, $\langle \psi \rangle$ and
$\gamma$. In the case of the type 1 decoherence processes we obtain
for $\tilde{\langle \psi
\rangle}$
\begin{equation}
\label{eq:last-field}
\tilde{\langle \psi \rangle}=\frac{\langle \psi \rangle}{2}+
\frac{\sqrt{2}\langle \psi \rangle}{4E_j}\sqrt{E_j^2-4\gamma_1^2
-\omega^2_n+\sqrt{-16\gamma^2_1E_j^2+(E_j^2+4\gamma^2_1+\omega^2_n)^2}},
\end{equation}
and for $\tilde{\epsilon_j}$ and $\tilde{\omega_n}$
\begin{align}
\label{eq:last-omega}
\tilde{\omega_n}&=\frac{\omega_n \tilde{\langle \psi\rangle}}
{2\tilde{\langle \psi \rangle} - \langle \psi \rangle}, \\
\label{eq:last-epsilon}
\tilde{\epsilon_j}&=\frac{\epsilon_j \tilde{\langle \psi
\rangle}}{\langle\psi \rangle},  
\end{align} 
while for the type 2 decoherence processes we have
\begin{multline}
\label{eq:last-field2}
\tilde{\langle \psi \rangle}=\frac{\langle \psi \rangle}{2}+ \\
\frac{\sqrt{2}\langle \psi \rangle}{4(\omega_n^2+g^2\langle \psi\rangle^2)}
\sqrt{E_j^2-2(\epsilon-\mu)^2+4\gamma_1^2-\omega^2_n+
\sqrt{16\gamma^2_2(\omega_n^2+g^2\langle \psi
\rangle^2)+(E_j^2-4\gamma^2_1+\omega^2_n)^2}}, 
\end{multline}
and 
\begin{align}
\label{eq:last-omega2}
\tilde{\omega_n}&=\frac{\omega_n \tilde{\langle \psi
\rangle}}{\langle\psi \rangle}, \\   
\label{eq:last-epsilon2}
\tilde{\epsilon_j}&=\frac{\epsilon_j \tilde{\langle \psi\rangle}}
{2\tilde{\langle \psi \rangle} - \langle \psi \rangle},
\end{align}
where for both cases
\begin{equation}
\label{eq:Equasip}
E_j=\sqrt{(\epsilon_j-\mu)^2+g^2\langle \psi \rangle^2}.
\end{equation}
Substituting the equations (\ref{eq:last-field}) -
(\ref{eq:last-epsilon}) or (\ref{eq:last-field2}) -
(\ref{eq:last-epsilon2}) into equation (\ref{eq:selfcons}),
summing over the Matsubara frequencies and using equation
(\ref{eq:field_pol}) we can determine the coherent polarisation
$\langle P \rangle$ and the coherent photon field $\langle \psi
\rangle$ as functions of the system parameters $\epsilon$, $\omega_c$,
decoherence parameter $\gamma_{1,2}$ and chemical potential $\mu$. The
chemical potential can be then obtained from the equation for the
number of excitations (\ref{eq:nex}). The integrals over the Matsubara
frequencies at zero temperature in the equations (\ref{eq:selfcons})
and (\ref{eq:selfcons}) as well as the determination of the chemical
potential from the equation (\ref{eq:nex}) have to be performed
numerically. 

Using the above method we could determine the ground state properties
of the system such as the ground state average of the coherent field, of
the coherent polarisation and of the inversion as well as the chemical
potential. Also the excitation spectrum, i.e the density of states, can
be obtained from the diagonal part of the Green function of a real
frequency. Considering the analytical continuation of $G(j,i\omega_n)
\to \bar G(j,\omega)$, where $\omega_n$ and $\omega$ are the Matsubara 
and the real frequencies respectively, and thus using the usual
substitution $i\omega_n \to \omega - i \delta$, we obtain the
relationship between the Green function $G(j,i \omega_n)$ and the
density of states $A(\omega)$
\begin{equation}
A(\omega)=\sum_j\lim_{\delta \to 0^+}ImG_{bb}(j,-\omega+i\delta+\mu),
\end{equation}
which in our case would take the form
\begin{equation}
\label{eq:spectrum}
A(\omega)=\sum_j Im \frac{\tilde{\omega}+(\tilde{\epsilon}_j-\mu)}
{\tilde{\omega}^2+(\tilde{\epsilon_j}-\mu)^2 +
g^2\tilde{\langle \psi \rangle}^2}.
\end{equation}
$\tilde{\omega}$, $\tilde{\epsilon_j}$ and $\tilde{\langle
\psi \rangle}$ are functions of $\omega$, $\epsilon_j$, $\langle
\psi \rangle$ and $\gamma$ which can be obtained from the equations
(\ref{eq:last-field}) - (\ref{eq:last-epsilon}) or
(\ref{eq:last-field2}) - (\ref{eq:last-epsilon2}) by the following
substitution $i\omega_n \to \omega - i \delta$. It can be shown by
analysing equation (\ref{eq:spectrum}) that the system of 
two-level oscillators with uniform energies, $\epsilon_j=\epsilon$, in
the presence of the type 1 processes has a gap, $\Delta$, in the
density of states of magnitude
\begin{equation}
\Delta=2\sqrt{(\epsilon-\mu)^2+g^2\langle \psi \rangle^2)}-4\gamma_1.
\label{eq:gap}
\end{equation}

The major difference between our calculations and the Abrikosov and
Gor'kov theory \cite{abrikosov-gorkov} is that we have two order
parameters connected through the chemical potential which needs to be
determined. We use a different form of the density of states for the
two-level oscillators than in their theory. Instead of a flat
distribution of energies from $-\infty$ to $+\infty$ used in the
Abrikosov and Gor'kov method we first perform the calculations for the
degenerate case where all two-level oscillators have the same energy
$\epsilon$ and then we use a realistic Gaussian distribution of
energies, present in the real microcavities. To account for these
differences we need to include the additional, third equation for
renormalised $\tilde{\epsilon}$ not present in the original Abrikosov and
Gor'kov method and the constraint equation for $n_{ex}$. In
the Abrikosov and Gor'kov theory they consider free propagating
electrons with momentum {\bf k} over which all the summations are
performed and the external impurities mix these {\bf k} states. In our
model of the localised two-level systems the summations are performed
over the sides where the two-level oscillators can be present and it is
assumed that the interactions with the environment do not mix these sides.

To perform the calculations we rescale the coherent fields by
$\sqrt{N}$ and consequently the inversion and the number of
excitations by $N$ introducing the excitation density
$\rho_{ex}=n_{ex}/N$. In this terminology the minimum $\rho_{ex}=-0.5$
corresponds to no photons and no electronic excitations in the
system. The condition $\rho_{ex}=0.5$ in the absence of photons would
correspond to all two-level oscillators in excited states, thus to the
maximum inversion.

\section{Results}

We calculate first the ground state coherent field $\langle \psi
\rangle$, the coherent polarisation $\langle P \rangle$, the inversion
and the chemical potential as functions of the decoherence strength
$\gamma$ and the excitation density $\rho_{ex}$. Then we study the
excitation spectrum of the system for different regimes. The ground
state properties and the excitation spectrum allow us to obtain a
phase diagram for different excitation densities and decoherence
strengths. We consider the influence of the type 1 and the type 2
decoherence processes and the inhomogeneous broadening of exciton
energies.

\subsection{Type 1 (Pair-Breaking) Decoherence Processes}
\label{type1}

We first consider in detail the influence of the type 1 decoherence
processes on the system of two-level oscillators with uniform energies
$\epsilon_j=\epsilon$.

\subsubsection{The Ground State - Coherent Fields}

\begin{figure}[htbp]
	\begin{center} 
	\leavevmode 
		\epsfxsize=8.8cm
		\epsfbox{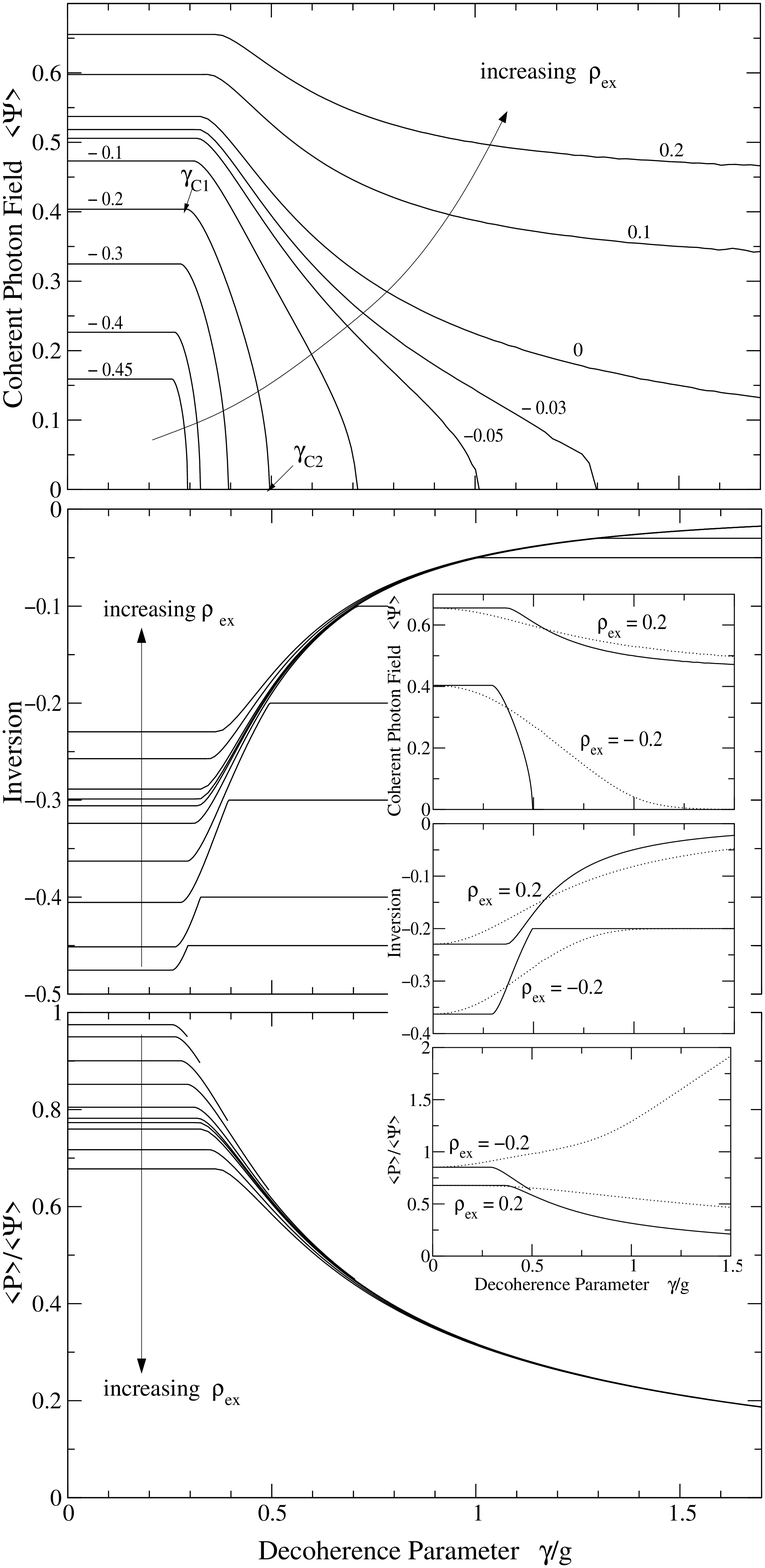} 
	\end{center} 
	\caption{Coherent photon field $\langle \psi \rangle$ (upper
         panel), inversion (middle panel) and ratio between
         coherent photon field and coherent polarisation $\langle P
         \rangle/\langle \psi \rangle$ (lower panel) as functions of
         the pair-breaking decoherence strength, $\gamma/g$ for
         different excitation densities, $\rho_{ex}$. Inset: Comparison
         between the influence of a pair-breaking (solid line) and
         a non-pair-breaking (dotted line) decoherence processes on
         $\langle \psi \rangle$ (upper panel), inversion (middle panel)
         and $\langle P \rangle/\langle \psi \rangle$ (lower panel)
         for two different values of $\rho_{ex}$. }
\label{fig:phot}
\end{figure}

To examine the ground state properties of the system we first
study the mean value of the annihilation operator of the field and the
polarisation. This mean is non-zero only in a coherent state. Figure
\ref{fig:phot} (upper panel) shows the behaviour of the coherent part of
the photon field $\langle \psi \rangle$ as the decoherence strength
$\gamma$ is changed for different excitation densities $\rho_{ex}$.
For small values of $\gamma/g$, up to some critical value
$\gamma_{C1}$, $\langle \psi\rangle$ is practically unchanged while
for $\gamma/g>\gamma_{C1}$ the coherent field is damped quite rapidly
with the increasing decoherence strength. This critical value of the
decoherence strength, $\gamma_{C1}$ is proportional to $\rho_{ex}$,
suggesting that for higher excitation densities the system is more
resistant to the decoherence processes. At low excitation densities,
where $\rho_{ex}<0$, there is a second critical value of the
decoherence strength, $\gamma_{C2}$, where both coherent fields are
sharply damped to zero. As the excitation density is increased,
precisely at $\rho_{ex}=0$, $\gamma_{C2}$ moves to infinity and it does
not exist for $\rho_{ex}>0$ - coherent fields although damped are
never completely suppressed.

The behaviour of the electronic inversion, given by the equation
(\ref{eq:inv}), is presented in Fig. \ref{fig:phot} (middle panel). In
this region of the decoherence strength where $\langle \psi \rangle$
is damped the inversion increases. At low excitation densities
($\rho_{ex}<0$) the inversion approaches $\rho_{ex}$ for
$\gamma/g=\gamma_{C2}$ and stays constant as $\gamma$ is further
increased. At high excitation densities ($\rho_{ex}>0$) the inversion
asymptotically approaches zero with increasing decoherence
strength.

The ratio of coherent polarisation to coherent field $\langle
P \rangle/\langle \psi \rangle$ is presented in Fig. \ref{fig:phot}
(lower panel). For an isolated system, where $\gamma=0$, this ratio
depends on the excitation density. The condensate becomes more
photon like as $\rho_{ex}$ is increased due to the phase space filling
effect. For finite $\gamma$, at a given excitation density, this ratio
decreases with increasing $\gamma$ meaning that the coherent
polarisation is more heavily damped than the coherent photon field by
the type 1 decoherence processes. At $\rho_{ex}<0$ this ratio becomes
undefined for $\gamma/g>\gamma_{C2}$ when both coherent fields vanish.

\subsubsection{Excitation Spectrum and the Phase Diagram}

\begin{figure}[htbp]
	\begin{center}
	\leavevmode
		\epsfxsize=16.2cm
		\epsfbox{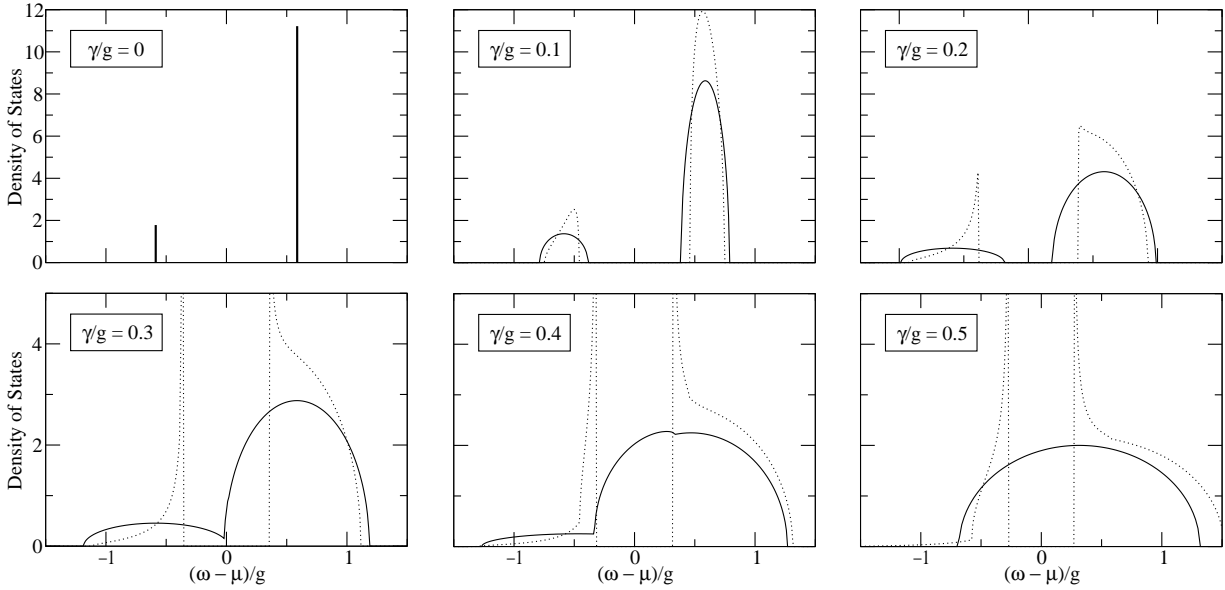}
	\end{center}
	\caption{Density of states for $\rho_{ex}=-0.2$ and different
  	decoherence strengths, $\gamma/g$ for a pair-breaking (solid
  	line) and a non-pair-breaking (dotted line) decoherence
  	processes.}
\label{fig:green}
\end{figure}

Now we examine the density of states for the system in different
regimes for different values of $\rho_{ex}$ and $\gamma$. In Fig.
\ref{fig:green} we present the density of states for $\rho_{ex}=-0.2$ 
and six different values of $\gamma$.
In the absence of decoherence (Fig. \ref{fig:green} a) we have two
sharp peaks at two quasi-particle energies, $\pm E$, given by 
equation (\ref{eq:Equasip}) for the uniform case ($E_j=E$). As
$\gamma$ increases these two peaks broaden, which causes the
decrease in the magnitude of the energy gap (Fig. \ref{fig:green} b
and c). The magnitude of the energy gap in the uniform case is equal
to $2E-4\gamma$, which is given in more detail in equation
(\ref{eq:gap}). Finally, precisely at $\gamma_{C1}$ (shown in Fig.
\ref{fig:phot}), these two broadened peaks join together and the gap
closes (Fig. \ref{fig:green} d). When the decoherence strength is
increased further (Fig. \ref{fig:green} e) these two peaks overlap
more and the shape of the gapless density of states changes. For
$\gamma/g>\gamma_{C2}$ the coherent fields are suppressed, thus Fig
\ref{fig:green} f shows the normal state density of states in the absence
of coherence. 

\begin{figure}[htbp]
	\begin{center}
	\leavevmode
		\epsfxsize=16.2cm
		\epsfbox{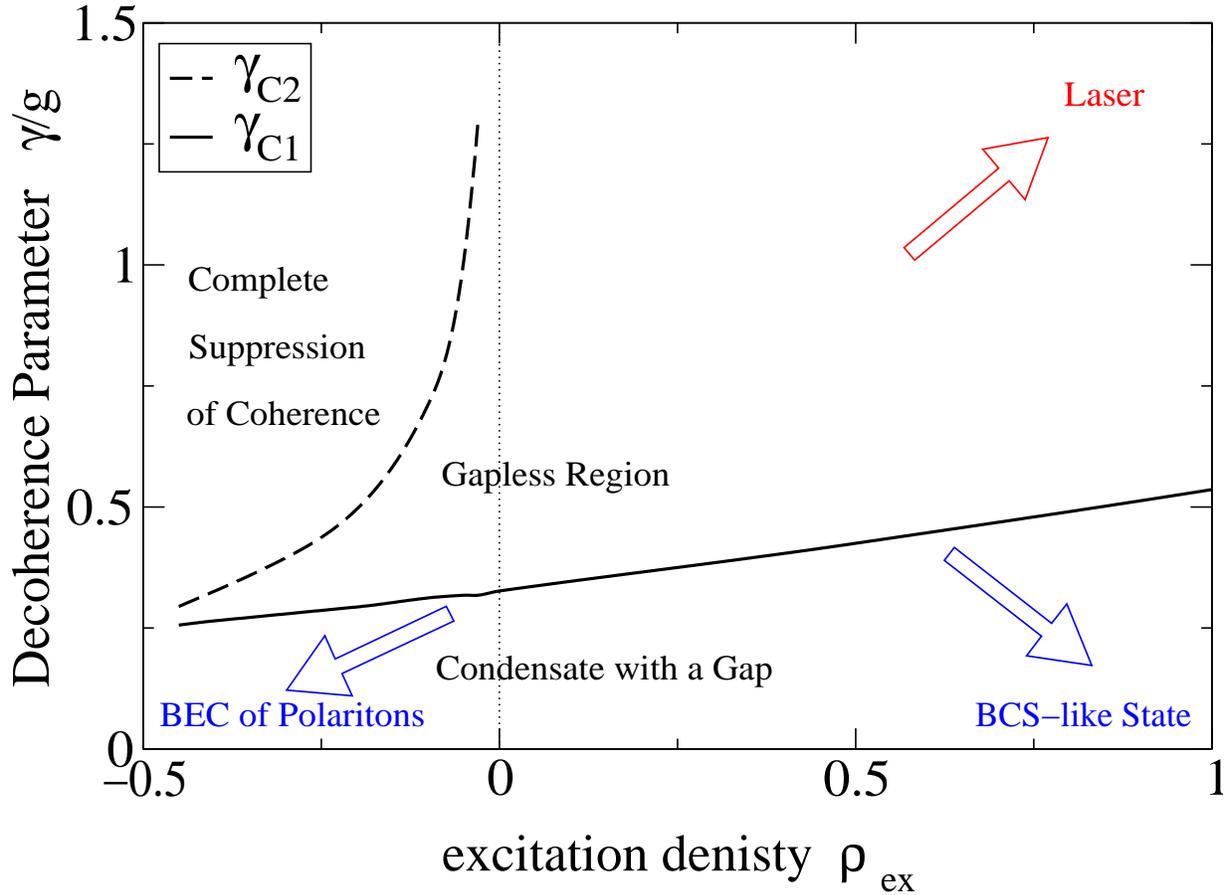}
	\end{center}
	\caption{Phase diagram. The phase boundaries $\gamma_{C1}$ and
	$\gamma_{C2}$ as marked in Fig \ref{fig:phot}.  }
\label{fig:phase}
\end{figure}

There are clearly three different phases depending on the decoherence
strength $\gamma$ and the excitations density $\rho_{ex}$. In the Fig.
\ref{fig:phase} we present a phase diagram for the system. 
The phase boundaries are defined by $\gamma_{C1}$ and $\gamma_{C2}$
for different values of $\rho_{ex}$. 

Below the solid line, for small decoherence, we have a phase in which
both coherent fields and the energy gap in the density of states are
present. In this region coherent fields are protected by the energy
gap and remain practically unchanged while the energy gap narrows as
the decoherence is increased. At low densities within this phase we
have a BEC of polaritons with the electronic and photonic parts
comparable in size. At high densities we have a BCS-type of condensate
with photon component increasing with the excitation density. Despite
the predominantly photon-like character of this phase at very high
densities, the coherence of the media is still large and this phase
can be distinguish from a laser by the presence of a gap in the
density of states (Fig \ref{fig:greenb} - right panel). The crossover
between high and low densities is a smooth evolution and there are no
rapid phase transitions.

Between the solid and dashed lines there is a phase where the coherent
fields are present without the energy gap in the density of
states. This phase exists for all values of the decoherence parameter
at the excitation densities $\rho_{ex} >0$. The coherent fields are no
longer protected by the gap and get damped as the decoherence is
increased. The coherent polarisation is more heavily damped than the
coherent field.  Within this phase we have at low densities a gapless,
light-matter condensate, analogous to gapless superconductivity whilst
at very high densities we have essentially the laser system.

Finally, above the dashed line there is a phase where the coherent
fields are completely suppressed.

\subsubsection{Crossover to Laser}

The features of the second phase (between the solid and the dashed
lines) in Fig. \ref{fig:phase} at high $\rho_{ex}$ are essentially the
same as those of the laser system. The laser operates in the regime of
a very strong decoherence, comparable with the light-matter interaction
itself. For a sufficiently large ratio of pumping and damping
processes, which sets the excitation density in the cavity, even in the
presence of such a large decoherence the laser action can be
observed. The coherent polarisation in a laser system is much more
heavily damped than the photon field and the gap in the density of
states is not observed. Thus the laser is a regime of our system for very
large $\rho_{ex}$ and $\gamma$.

Laser theories, due to the approximations on which they are based, can
only be valid in a regime where the gap in the density of states is
suppressed and thus for a large decoherence. At the time when these
theories were proposed they were deemed sufficient as most lasers
operate in such a regime. Miniaturisation and improvements in the
quality of optical cavities in recent years can lead to a large
suppression of decoherence in a laser media. For small decoherence and
very small pumping in comparison to decay processes, when $\rho_{ex}$
is much smaller then $0$, the laser theories would predict a lack of
coherence while the real ground state of the system would be a more
matter-like condensate. Thus an extension of laser theories to account
for the gap in the excitation spectrum and coherence in a media is
necessary.

When both the gap and the coherence in a media are taken into account,
in contrast to a traditional laser, the coherent photon field can be
present without the population inversion in the media. The polariton
condensate is thus an example of a laser without inversion.

It has to be pointed out that the distinction between a laser and a
Bose condensate of polaritons is not a particularly clear one. In the
laser, coherence in the media (manifested by the coherent
polarisation) although small, is not completely suppressed so the
laser can be seen as a gapless condensate with a more photon-like
character. One of the possible distinctions between BEC and laser
could be an existence of an energy gap in the excitation spectrum.

\subsection{Type 2 (Non-Pair-Breaking) Decoherence Processes}
\label{type2}

We have also studied the type 2 decoherence processes, analogous to
the non-magnetic impurities in superconductors, which act in an
exactly the same way on the upper and the lower levels of the
two-level oscillator (the second term in the expression
\ref{eq:Hmag}). We found that these processes give rise to the
broadening of energies of the two-level system and do not have any
pair-breaking effects. In the Abrikosov and Gor'kov theory, due to the
flat density of sates used in the calculations, the non-magnetic
impurities do not influence the superconducting state at all. In the
case of uniform or realistic, Gaussian distributed energies of the
two level oscillators the type 2 processes have some quantitative
influence on the coherent fields and the gap but cannot cause any
phase transitions. We now perform the self-consistent method described
in Section \ref{method} for the type 2 decoherence processes. The
results are shown as dotted lines in onset of Fig. \ref{fig:phot} and
in Fig. \ref{fig:green}.

\subsubsection{The Ground State - Coherent Fields}

In the inset of Fig. \ref{fig:phot} we present for comparison the
coherent photon field (upper panel), the inversion (middle panel), and
the ratio between the coherent polarisation and the coherent photon
field (lower panel) in the presence of the type 1 (solid line) and the
type 2 (dotted line) decoherence processes for two excitation
densities $\rho_{ex}=-0.2$ and $\rho_{ex}=0.2$. It can be noticed that
the sharp phase transitions at $\gamma_{C1}$ and $\gamma_{C2}$
discussed for the type 1 decoherence processes are not present for the
case of the type 2 processes. With the increase of the decoherence
strength from zero the coherent fields are slightly damped and they
slowly decrease, asymptotically approaching zero at low excitation
densities ($\rho_{ex}<0$) or a constant value at high
densities ($\rho_{ex}>0$). Although both coherent fields are damped the
behaviour of their ratio strongly depends on the excitation
density. Since the type 2 processes give rise to the broadening of
energies, the behaviour of the ratio $\langle P \rangle/\langle \psi
\rangle$ depends on the position of the chemical potential with
respect to the energy distribution. Thus the type 2 processes can make
the condensate more photon or more exciton like depending on the
parameters of the system. The two different cases are presented in 
Fig. \ref{fig:phot}.

\subsubsection{Excitation Spectrum}

The density of states in the presence of type 2 processes is shown
in Fig \ref{fig:green} (dotted lines) for the same parameters as
were used in Section \ref{type1} for studying the influence of 
type 1 decoherence processes. It can be noticed that, although the two
quasiparticle peaks get very broad, the gap is only slightly affected
by the type 2 processes and is always present. Even for much larger
values of $\gamma$ than presented in Fig \ref{fig:green} the gap is
not suppressed. The gap in the density of states is present until the
coherent fields get completely suppressed. At high excitation densities
($\rho_{ex}>0$) coherent fields are always present and thus the
density of states will have a gap for all values of the decoherence
strength.

Despite a different physical origin the type 2 processes give similar
effect as the inhomogeneous broadening of energy levels in the case of
an isolated system (see \cite{paul,paul-phd}).

\subsection{Inhomogeneous Broadening of Energies}

Finally, we study the influence of the type 1, pair-breaking processes
on the system of realistic, inhomogeneously broadened two-level
oscillators.  We replace the summations over sites with integrals over
the energy distribution. We assume this distribution to be a Gaussian
with mean $\epsilon_0$ and variance $\sigma $.

\begin{figure}[htbp]
	\begin{center}
	\leavevmode
		\epsfxsize=16.2cm
		\epsfbox{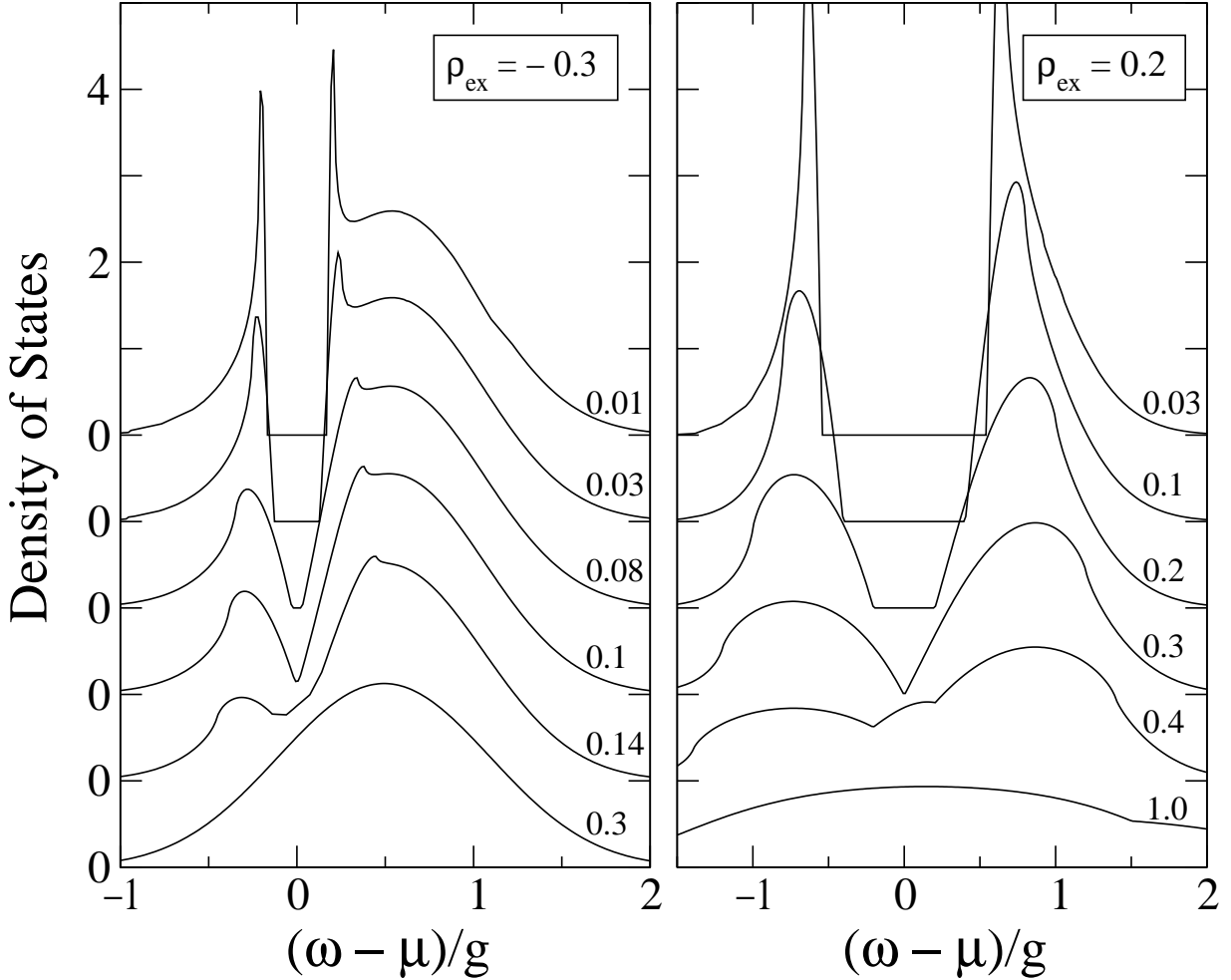}
	\end{center}

	\caption{Density of states for the Gaussian broaden case
	with $\sigma=0.5$ for different values of the pair-breaking
	decoherence strength at $\rho_{ex}=-0.3$ (left panel) and
	$\rho_{ex}=-0.2$ (right panel). }

\label{fig:greenb}
\end{figure}

Our results show that the Gaussian broadening of energies does not add
any qualitative difference to the uniform case. The coherent fields
are, as expected, slightly smaller than in the degenerate case but all
the regimes and transitions described in the Section \ref{type1} are
also present here. The critical values of decoherence strength,
$\gamma_{C1}$ and $\gamma_{C2}$ are slightly smaller and the energy
gap narrower than in system without energy broadening for the same
parameters. Fig \ref{fig:greenb} shows the density of states for the
system of the inhomogeneously broadened two-level oscillators with
standard deviation $\sigma=0.5g$ for different values of the type 1
decoherence strength at two different excitation densities.
The broadening of the density of states and the suppression of the
energy gap can be observed as $\gamma$ is increased. The last curve in
the Fig \ref{fig:greenb} (left panel) shows the normal state density
of states where the coherent fields are suppressed. In the Fig
\ref{fig:greenb} (right panel) where the coherent fields are present
for all the values of $\gamma$ the last curve has no signs of the gap.

As we have pointed out in Section \ref{type2} the type 2 decoherence
processes give similar effects as the inhomogeneous broadening of
energy levels. The difference is that the density of states in the
presence of the type 2 processes has sharp boundaries. This is a
result of the method which is used to perform the calculations. It
can be noticed that if the bath operators for the type 2 processes,
$c^{\theta}$'s in expression (\ref{eq:Hmag}), were just numbers, the
second term in expression (\ref{eq:Hmag}) could had been included into
the first term in equation (\ref{eq:H0}) and would had just given a
random $\epsilon_i$. In our calculations the $c^{\theta}$'s are
operators and we use a self-consistent Born approximation, which is
not exact, and makes errors in the tail of the distribution. Thus,
because of the way the calculations are done, the density of states
produced always has sharp boundaries.

\section{Conclusions}

In this work we have studied the influence of two kinds of decoherence
processes and energy broadening on the system of interacting two-level
oscillators (atoms or excitons) with the photon mode in the cavity. We
have noticed that the widely used Langevin equations with the decay
constants independent of frequency cannot be used for a small
decoherence when gap in the density of states is present.  We propose
a self-consistent method similar to the Abrikosov and Gor'kov theory of
gapless superconductivity \cite{abrikosov-gorkov} which allow us to
study all regimes of the system as the decoherence is changed from
zero to large values and for a wide range of excitation densities. At
low excitations and small decoherence we obtain a coupled matter-light
condensate with a gap in the density of states. As the decoherence is
increased at low excitation densities we obtain a gapless condensate
and finally a complete suppression of coherence. With an increase in
the excitation density the condensate becomes more photon like, but
a large gap in the density of states is present for a small
decoherence strength. In the limit of a very large excitation density
and a large decoherence we reach a gapless, laser-like regime in which
the coherent polarisation is very small in comparison with the
coherent photon field.

We have shown that the polariton condensate, even in the presence of
the interactions with an environment, is protected by the energy gap
and thus should be observable if the decoherence is not too
large. This gap increases with an increase in the excitation density
and thus, in contrast to excitonic insulator, polariton condensate
should be present at high densities.

Our results suggest that, in contrast to lasers, a coherent light can
be generated without a population inversion. This work generalised the
existing laser theories to include the gap in the excitation spectrum
caused by the coherence in a media in the low decoherence regime.

We have studied separately the two types of decoherence processes in
order to understand their different influence on the system. Although
in the real cavities both decoherence processes and inhomogeneous
broadening of energies will be present, only the pair-breaking, type
1, processes are responsible for the phase transitions and the
suppression of the gap in the density of states. The type 2 processes
and the inhomogeneous broadening of energies cannot on its own lead
to the phase transitions. However they would make the amplitude of the
coherent fields smaller and thus, in the presence of the pair-breaking
processes, would cause these transitions to be present at smaller
decoherence strength. For the quantitative analysis and the comparison
with experiments all the decoherence processes should be included.

We have treated all the decoherence processes in a phenomenological
way to study their qualitative influence on the general system. In the
case of a particular example, when the origin and the density of states
for the environment is known, it would be possible to perform more
detailed calculations. In general the bath propagators in the
equations (\ref{eq:selfener1}) and (\ref{eq:selfener2}) would be
frequency dependent and thus $\langle \psi \rangle$ would depend on
the frequency as well. This frequency dependence of the coherent
field, as in an Eliashberg theory of superconductivity
\cite{eliash,scalapino}, would have to be included in the calculations.

In this work we have studied the influence of the decoherence on an
equilibrium, incoherently pumped system. However, in the same way it
would be possible to examine the influence of an environment on
coherently driven condensates discussed in the Section \ref{driven}.
The coherent photon field in the driven system is an external, fixed,
parameter and not a self-consistent field satisfying equation
(\ref{eq:selfcons}). In contrast the gap in the density of states,
proportional to the coherent field amplitude, is present exactly as
for an equilibrium condensate. Thus the decoherence processes in the
coherently driven system should be treated in the same way as for an
equilibrium condensate described in this Chapter. In a recent paper
\cite{wrong} authors have ignored this fact and performed the Langevin
equation method with the constant decay rates to study the stability
of the driven excitonic condensate concluding that the condensed phase
will be destroyed due to an arbitrarily small decoherence. In our
opinion this is not physical and results only from the performed
approximations.

\chapter{Conclusions and Future Directions}
\label{concl}

In this work we have developed a method to include general decoherence
effects in the systems with an energy gap in the excitation
spectrum. We have shown that the widely used Langevin equations with
the constant, frequency independent decay rates are not valid in
the regime where the gap is present. The decoherence has to be treated
in a self-consistent way similar to the Abrikosov and Gor'kov theory
of gapless superconductivity. We studied the general form of the
interactions, introducing pair-breaking and non-pair-breaking
processes and thus our theory is applicable to a wide range of
systems. For more detailed results, the particular origin of the
interactions and thus the particular density of states for the baths
have to be taken into account.

This method allows us to study the stability of the polariton
condensate.  For low decoherence strength the condensate is
protected by the gap in the density of states. This gap is
proportional to the amplitude of the coherent photon field and thus is
larger for higher densities. As the decoherence is increased the gap
gets smaller and finally is completely suppressed.

Experiments on cavity polaritons which report the stimulated
scattering into the lower polariton branch \cite{dang} -
\cite{angle-cw-stimscat} (see Section \ref{polariton}) are performed
at low densities to avoid the fermionic phase space filling effect, so
that two clear polariton peaks could be seen.  For such densities
the gap in the excitation spectrum would be very small and easily
suppressed by the decoherence processes in the sample. Indeed such a gap
is not observed in the photoluminescence spectrum. The attempt to
increase the density of polaritons in these experiments results not in
the formation of the condensate but in a switching into the
weak-coupling regime and lasing.

This is not surprising since the increase in the density of polaritons
in these experiments is obtained by increasing the pumping of
excitons. As we discussed in Chapter \ref{abricosov}, the incoherent
pumping is a pair-breaking decoherence process and thus the increase
in the pumping intensity results in the increase of not only the
density of excitons but also the decoherence strength. As shown
by Eastham and Littlewood \cite{paul,paul-phd}, the fermionic
structure of excitons does not prevent condensation, even at very high
densities. Thus, in the current experiments it is not a phase-space
filling effect which lead to a laser as the pump intensity is
increased, but the increase in the decoherence strength. If the
polariton condensate is ever to be observed an increase in density
without an increase in decoherence is necessary.

The density of excitons is proportional to the ratio between the
pumping and the cavity decay rates. This ratio can be large even for
small values of both rates. Thus, in order to achieve high
densities without large decoherence the cavity decay rate must be
small, which suggests that experiments should concentrate on
improving the quality of the mirrors in microcavity systems.

The localised and tightly bound excitons, like excitons in disordered
quantum wells or Frenkel-type excitons in organic compands, seem to be
better candidates for observing the polariton condensate than the
high-quality GaAs quantum wells with weakly bound, delocalised
excitons. Static disorder is not a pair-breaking effect and would
have a weaker influence on the condensate than screening and
ionisation in the case of delocalised or weakly bound excitons.

Our method allows us to study different regimes as the decoherence and
the excitation density is changed and thus to examine the crossover
from a polariton condensate to a laser. The laser emerges from the
polariton condensate at high densities when the gap in the density of
states closes for large decoherence and thus is analogous to a gapless
superconductor. Our work generalises the existing laser theories to
include coherence effects in the media. When this coherence is
included the generation of a coherent photon field without 
population inversion becomes possible.

In the case of coherently driven systems, in which a gap in the
density of states would be present, the decoherence processes have to
be also treated in a self-consistent way, similar to that described in
this work.

Our method is not limited to semiconductor excitons and can be
applied to any bosonic excitations which can be described as two-level
oscillators, like two electronic levels in atoms or molecules.

There are many directions in which our work could be further
extended. In the next Section we will review this possible directions
and indicate the work being undertaken at the moment.

\section{Future Work}

\subsection{Finite Temperatures}

The results presented in this work are performed at zero temperature
for which the summations over the Matsubara frequencies in equations
(\ref{eq:selfcons}) and (\ref{eq:inv}) become integrals and are
performed numerically. The extension to finite temperatures is
straightforward. At finite temperatures the summations over a discrete
$\omega_n$, which in the case of fermions would be
$\omega_n=(2l+1)\pi/\beta$ can be performed using the standard
techniques \cite{ben,green} and thus finite temperatures could be
easily studied.

\subsection{Non-Equilibrium}

A much more important extension that the finite temperature case is the
problem of non-equilibrium systems. For a system with very strong
pumping and decay processes the thermal distribution of the relevant
quasiparticles cannot be assumed. It would be of a great interest
to examine the influence of non-equilibrium on the behaviour of this
system. This would also allow us to directly compare our theory to
experiments on stimulated scattering of polaritons which are mainly
out of equilibrium. We would also be able to directly apply our theory
to coherently driven systems which are also out of thermal
equilibrium. This work is currently in progress.

In this work we included the pair-breaking effect of pumping by
studying the processes described by the first term of expression
(\ref{eq:Hmag}), which has the same decoherence effect as pumping
but do not cause the system to be in a non-equilibrium state. The ratio
between the pump and the decay rates was introduced in terms of the
excitation density $\rho_{ex}$. To study the influence of the system being
out of equilibrium we fully include the decay of the cavity field and the
pumping (the first, second and third terms in equation (\ref{eq:HSB}))
which would
cause a flow of energy through the system. We do not need to consider
the constraint (\ref{eq:nex}) now, as the density of the system would
get fixed in a steady state by the relative strength of the pumping
and the decay processes. The photon field $\langle \psi \rangle$ in
this case becomes a function of frequency in a similar way to an
Eliashberg theory of strong coupling superconductors
\cite{eliash,scalapino}. All the external decoherence processes need
to be included in a similar way as presented in this work so that the
gap in the density of states is taken into account. The
non-equilibrium Keldysh techniques need to be applied to obtain the
distribution of particles in the steady state. Thus, using the analogy
to superconductivity, we need to include the decoherence processes as
in the Abrikosov and Gor'kov theory of gapless superconductivity, the
frequency dependent photon field as in an Eliashberg theory of strong
coupling superconductor and the flow of the energy through the system in
a non-equilibrium state in the same framework. It is not surprising that
it is very difficult, if not impossible, to treat such a complex
problem within the Green's functions techniques presented in this
work. We think that the more powerful field theoretical methods, like
a non-linear sigma model, would be more suitable here. The
implementation of this method for our model is still in progress.

\subsection{Microscopic Details of the Decoherence Processes}
\label{future-decoh}

If the microscopic origins of the decoherence processes are known for a
particular experimentally studied system, the theory could be extended
to include this information. The detailed account of the coupling
constants and the density of states for the environment can be easily
included within this framework. In general the baths propagators in
the equations (\ref{eq:selfener1}) and (\ref{eq:selfener2}) would be
frequency dependent and thus $\langle \psi \rangle$ would depend on
frequency as in the Eliashberg theory of gapless superconductors
\cite{eliash,scalapino}. The phenomenological constants $\gamma_1$ and
$\gamma_2$ can thus be, in principle, obtained from this microscopic
calculation for a particular system.

\subsection{Generalisation of the Model}
\label{future-gen}

Finally, we can generalise the model itself to account for a different
level of disorder and a binding energy of excitons in a material. As
we discussed in the Section \ref{paul}, the samples are characterised
by a different level of fluctuations in the thickness of the quantum well,
alloying concentration or presence of impurities. The spatial extent
of the excitons also depends on the materials being used and vary from
tightly bound, Frenkel-type excitons to weakly bound, extended Wannier
excitons.  Although in light of recent experiments \cite{gammon}
- \cite{bonadeo}, the localised exciton picture seems the most relevant
the theory could be extended to account for different situations.

In the absence of disorder, when the exciton centre of mass wavefunction
would propagate through the sample, the direct Coulomb interactions
could be incorporated at a mean-field level in a similar way as in the
case of a coherently driven excitonic insulator \cite{schmitt}. As we
discussed in the Section \ref{model}, adding the direct Coulomb
interactions between excitons within a mean-field would only give
small corrections to the theory presented in this work. At high
densities the Coulomb interaction is very small in comparison to the
dominant dipole interaction while at low densities the dominant part
of the Coulomb interaction is the one between an electron and a hole
within the same exciton, which is included in our model. But for 
qualitative analysis of a particular media this correction could be
included.

For materials with Frenkel excitons the theory could be extended to
include several localised exciton states on each site rather that
just the one assumed in this work. In the case of a narrow disorder
quantum well such a generalisation would also be possible in principle
but the electronic structure of these systems is not yet well studied
and thus quantitative description would be difficult.

\section{Summary}

We have studied the equilibrium Bose condensation of cavity polaritons
in the presence of decoherence. We have shown
\begin{itemize}
\item That the widely used Langevin equations with the constant,
frequency independent decay rates are not valid for systems in
which the gap in the density of states is present.
\item That in the regime of weak decoherence the decoherence
processes have to be treated self-consistently, in a way that the gap
in the density of states is taken into account.
\end{itemize}
We use self-consistent Green's function techniques, similar to the
Abrikosov and Gor'kov theory of gapless superconductivity
\cite{abrikosov-gorkov}, to  study different regimes in microcavity as
the decoherence strength and the excitation density is changed. We have
shown
\begin{itemize}
\item That at small decoherence the polariton condensate is protected
by the energy gap in the excitation spectrum. The gap is proportional to
the coherent field amplitude and thus the excitation density, so
the condensate is more robust at high densities.
\item That this gap gets smaller and eventually is completely
suppressed as the decoherence is increased.
\item That there is a regime, analogous to the gapless superconductor,
when the coherent fields are present without an energy gap. This
regime, at very high excitation densities, have all the features of a
photon laser. 
\end{itemize}
We study the influence of the two different types of processes, a
pair-breaking and a non-pair-breaking ones as well as the inhomogeneous
broadening of the energies. We have shown 
\begin{itemize}
\item That only the type 1, pair-breaking processes can lead to 
phase transitions.
\item That the type 2, non-pair-breaking processes and the
inhomogeneous broadening of energies can give only quantitative
influence to the behaviour of the system and can not prevent the
condensation. 
\end{itemize}
We have
\begin{itemize}
\item Studied the phase diagram of the system given by the Hamiltonian
(\ref{eq:H}) for different values of the decoherence strength and the
excitation densities.
\item Established the crossover between an isolated polariton
condensate and a photon laser as the decoherence strength is increased.
\item Generalised the existing laser theories to include coherence
effects in the media and the energy gap in the excitation spectrum.
\item Shown that, unlike in the traditional laser, the coherent photon
field can be present without a population inversion when the
coherence effects in the media are taken into account.
\end{itemize}

\addcontentsline{toc}{chapter}{Bibliography}
\bibliographystyle{acm}

\begin{thebibliography}{100}

\bibitem{bec}
A.~Griffin, D.~W. Snoke, and S.~Stringari, editors, {\em Bose-Einstein
Condensation}, Cambridge University Press, Cambridge, U.K., 1995.

\bibitem{nozieresbec}
P. Nozi{\`{e}}res, in {\em Bose-Einstein Condensation}, edited by
  A. Griffin, D.~W. Snoke, and A. Stringari (Cambridge University
  Press, Cambridge, U.K., 1995), pp.\ 15--30.

\bibitem{keldysh1}
L. V. Keldysh and Y. V. Kopaev, { \em Fiz. Tverd. Tela} {\bf 6}, 2791
(1964), [Sov. Phys. Solid State {\bf 6}, 2219, (1965)].

\bibitem{keldysh2}
L. V. Keldysh and A. N. Kozlov, {\em Zh. Eksp. Teor. Fiz.}  
{\bf 54}, 978 (1968), [Sov. Phys. JETP {\bf 27}, 521 (1968)].

\bibitem{nozieresex1}
C. Comte and P. Nozi{\`{e}}res, {\em J. Phys.} (Paris) {\bf 43}, 1069
(1982).

\bibitem{moscalenko}
S.~A. Moskalenko and D.~W. Snoke, {\em Bose-Einstein Condensation of
Excitons and Biexcitons}, CUP, Cambridge, U.K., 2000.

\bibitem{zittart}
J.~Zittartz, {\em Phys. Rev.} {\bf 164}, 575 (1967)

\bibitem{abrikosov-gorkov}
A.~A.~Abrikosov, L.~P.~Gor'kov, JETP {\bf 12}, 1243 (1960)

\bibitem{lin}
J. L. Lin, J. P. Wolfe,  {\em Phys. Rev. Lett.} {\bf 71},
1223 (1993).

\bibitem{mys1}
D. W. Snoke, J. P. Wolfe, A. Mysyrowicz, {\em Phys. Rev. B} {\bf 41},
11171 (1990).

\bibitem{mys2}
A. Mysyrowicz, D. W. Snoke, J. P. Wolfe, {\em Phys. Status. Solidi B}
{\bf 159}, 387 (1990).

\bibitem{mys3}
E. Fortin, S. Fafard A. Mysyrowicz, {\em Phys. Rev. Lett.} {\bf 70},
3951 (1993).

\bibitem{mys4}
A. Mysyrowicz, E. Fortin, E. Benson, S. Fafard, and E. Hanamura, {\em
Solid State Commun.} {\bf 92}, 957 (1994).

\bibitem{mys5}
E. Benson, E. Fortin, A. Mysyrowicz, {\em Phys. Status. Solidi B}
{\bf 191}, 345 (1995).

\bibitem{wind1}
A. E. Bulatov, S. G. Tikhodeev, {\em Phys. Rev. B} {\bf 46},
15058 (1992).

\bibitem{wind2}
S. G. Tikhodeev,  {\em Phys. Rev. Lett.} {\bf 78},
3225 (1997).

\bibitem{wind3}
S. G. Tikhodeev, G. A. Kopelevich, N. A. Gippus {\em
Phys. Status. Solidi B} {\bf 206}, 45 (1998).

\bibitem{wolfe}
K. E. O'Hara, J. P. Wolfe, Phys. Rev. B 62, 12909 (2000)

\bibitem{butov1}
L. V. Butov, A. Zrenner, G. Abstreiter, G. B\"{o}hm, G. Weimann, {\em
Phys. Rev. Lett.} {\bf 73}, 304 (1994).

\bibitem{butov2}
L. V. Butov {\em et al.}, {\em Surf. Sci.} { \bf 361/362}, 243 (1996)

\bibitem{butov3}
L. V. Butov, A. L. Ivanov, A. Imamoglu, P. B. Littlewood,
A. A. Shashkin, V. T. Dolgopolov, K. L. Campman, A. C. Gossard, 
{\em Phys. Rev. Lett.} {\bf 86}, 5608 (2001).

\bibitem{gelesin}
V.~M. Galitskii, S.~P. Goreslavskii, and V.~F. Elesin, { \em
Zh. Eksp. Teor. Fiz.}  {\bf 30}, 207 (1969), [Sov. Phys. JETP {\bf
30}(1), 117--122].

\bibitem{elesincop}
V. F. Elesin, Y. V. Kopaev, {\em Zh. Eksp. Teor. Fiz.} {\bf 63}, 1447
(1972), [Sov. Phys. JETP {\bf 36}(4), 767--770].

\bibitem{schmitt}
S. Schmitt-Rink, D.~S. Chemla, and H. Haug, { \em Phys. Rev. B} {\bf
37}, 941 (1988).

\bibitem{hopfpol}
J.~J. Hopfield, {\em Phys. Rev.} {\bf 112}(5), 1555 (1958).


\bibitem{atompol} 
M.~G. Raizen, R.~J. Thompson, R.~J. Breccha, H.~J. Kimble, and H.~J.  
Carmichael, {\em Phys. Rev. Lett.} {\bf 63}, 240 (1989).

\bibitem{cavpol}
C. Weisbuch, M. Nishioka, A.Ishikawa, and Y.Arakawa, {\em
Phys. Rev. Lett.} {\bf 69}, 3314 (1992).

\bibitem{bulkcavpol} 
A. Tedicucci, Y. Chen, V. Pellegrini, M. Borger, L. Sorba, F. Beltram,
and F. Bassani, {\em Phys. Rev. Lett.} {\bf 75}, 3906 (1995).

\bibitem{organicpol1} 
D.~G. Lidzey, D.~D.~C. Bradley, M.~S. Skolnick, T.~Virgili, S.~Walker,
and D.~M. Whittaker, {\em Nature} {\bf 395}, 53 (1998).

\bibitem{organicpol2}
D.~G. Lidzey, D.~D.~C. Bradley, T.~Virgili, A.~Armitage,
M.~S. Skolnick, and S.~Walker, {\em Phys. Rev. Lett.} {\bf 82}, 3316 (1999).

\bibitem{chargedpol}
R. Rapaport, R. Harel, E. Cohen, A. Ron, and E. Linder,
{\em Phys. Rev. Lett.} {\bf 84}, 1607 (2000).


\bibitem{dang}
L.~S. Dang, D.~Heger, R.~Andr{\'{e}}, F.~B{\oe}uf, and R.~Romestain,
{ \em Phys. Rev. Lett.} {\bf 81}, 3920 (1998).

\bibitem{senbloch}
P. Senellart and J. Bloch, { \em Phys. Rev. Lett.} {\bf 82},  1233  (1999).

\bibitem{bulkboser}
V. Pellegrini, R. Colombelli, L. Sorba, and F. Beltram, { \em Phys. Rev. B}
{\bf 59}, 10059 (1999).

\bibitem{baumberg} 
J. J. Baumberg, P. G. Savvidis, R. M. Stevenson, A. I. Tartakovskii,
M. S. Skolnick, D. M. Whittaker, and J. S. Roberts, { \em
Phys. Rev. B} {\bf 62}, R16247 (2000).

\bibitem{angleboser}
P.~G. Savvidis, J.~J. Baumberg, R.~M. Stevenson, M.~S. Skolnick, D.~M.
Whittaker, and J.~S. Roberts, { \em Phys. Rev. Lett.} {\bf 84}, 1547
(2000).

\bibitem{angle-cw-stimscat} 
R.~M. Stevenson, V.~N. Astratov, M.~S. Skolnick, D.~M. Whittaker,
E.~Emam-Ismail, A.~I. Tartakovskii, P.~G. Savvidis, J.~J. Baumberg,
and J.~S.  Roberts, {\em Phys. Rev. Lett.} {\bf 85}, 3680 (2000).


\bibitem{paul-phd}
P.~R. Eastham, {\em Ph.D. thesis}, Cambridge University, 2000. 


\bibitem{savona}
V.~Savona, S.~Haacke, and B.~Deveaud, {\em Phys. Rev. Lett.} {\bf
84}(1), 183--186 (2000).


\bibitem{paul} 
P.~R. Eastham and P.~B. Littlewood, {\em Solid State Commun.} {\bf
116}, 357 (2000).


\bibitem{gammon}
D. Gammon, E.~S. Snow, B.~V. Shanabrook, D.~S. Katzer, and D.~Park,
{\em Phys. Rev. Lett.} {\bf 76},  3005  (1996).

\bibitem{hegarty1}
J.~Hegarty, M.~D. Sturge, C.~Weisbuch, A.~C. Gossard, and W.~Wiegmann,
{\em Phys. Rev. Lett.} {\bf 49}(13), 930--932 (1982).

\bibitem{hegarty2}
J.~Hegarty, L.~Goldner, and M.~D. Sturge,
{\em Phys. Rev. Lett.} {\bf 30}(12), 7346--7348 (1984).

\bibitem{hess}
H.~F. Hess, E.~Betzig, T.~D. Harris, L.~N. Pfeiffer, and K.~W. West,
{ \em Science} {\bf 264}, 1740 (1994).

\bibitem{bonadeo}
N.~H. Bonadeo, G.~Chen, D.~Gammon, D.~S. Katzer, D.~Park, and
D.~G. Steel, {\em Phys. Rev. Lett.} {\bf 81}, 2759 (1998).

\bibitem{sculzub} 
For the quantum theory of the laser developed by
Haken, Risken, Lax, Louisell, Scully and Lamb see M.~O. Scully and
M.~S. Zubairy, {\em Quantum Optics} (Cambridge University Press,
Cambridge, U.K., 1997).

\bibitem{haken1}
H.~Haken, {\em Rev. Mod. Phys.} {\bf 47}, 67 (1975).

\bibitem{haken2}
H.~Haken, {\em Laser Theory}, Springer-Verlag 1984.

\bibitem{wrong}
K. Hannewald, S Glutsch and F. Bechstedt, {\em
J. Phys. Condens. Matter} {\bf 13}, 275 (2001)  

\bibitem{skalski}
S.~Skalski, O.~Betbeder-Matibet, P.~R.~Weiss { \em Phys. Rev.} {\bf
136}, A 1500 (1964)

\bibitem{green} 
G. Rickayzen, {\em Green's Functions and Condensed Matter}
(Academic Press Inc., London, 1987).

\bibitem{eliash}
G. M. Eliashberg, {\em Zh. Eksperim. i Teor. Fiz.} {\bf 38}, 966
(1960), [Sov. Phys. JETP {\bf 11}, 696 (1960)].

\bibitem{scalapino}
D.~J.~Scalapino, in {\em Superconductivity}, edited by R.~D. Parks,
volume~1, chapter~10, pages 449--560, Marcel Dekker, Inc., New York,
1969.

\bibitem{ben}
B.~D. Simons, ``Concepts in Theoretical Physics'',
Notes on Lectures given at Cambridge University, presently available
from http://www.tcm.phy.cam.ac.uk/$^\sim$bds10/.

\bibitem{keldyshbec}
L.~V. Keldysh, in {\em Bose-Einstein Condensation}, edited by
A. Griffin, D.~W. Snoke, and A. Stringari (Cambridge University Press,
Cambridge, U.K., 1995).

\bibitem{dickemodel}
R.~H. Dicke, {\em Phys. Rev.} {\bf 93},  99  (1954).

\end{thebibliography}

\end{document}